\documentclass[twocolumn]{aastex63}
\usepackage{amsmath}
\usepackage{enumerate}
\usepackage{enumitem}   
\usepackage{color}
\usepackage{subfigure}
\usepackage{color}
\usepackage{courier}
\usepackage{mathtools}

\newcommand{\GDSFR}{$\Delta$$log$SFR}
\newcommand{\SDSFR}{$\Delta$$log$sSFR($r$)}
\revised{}
\accepted{}
\submitjournal{}

\shorttitle{Spatially-resolved SFR in Interacting Galaxies}
\shortauthors{Pan et al.}
\begin{document}

\title{SDSS-IV MaNGA: Spatial Evolution of Star Formation Triggered by Galaxy Interactions}

\email{hapan@asiaa.sinica.edu.tw}

\author{Hsi-An Pan}
\affil{Institute of Astronomy and Astrophysics, Academia Sinica, 11F of AS/NTU Astronomy-Mathematics Building, No.1, Sec. 4, Roosevelt Rd, Taipei 10617, Taiwan}

\author{Lihwai Lin}
\affil{Institute of Astronomy and Astrophysics, Academia Sinica, 11F of AS/NTU Astronomy-Mathematics Building, No.1, Sec. 4, Roosevelt Rd, Taipei 10617, Taiwan}

\author{Bau-Ching Hsieh}
\affil{Institute of Astronomy and Astrophysics, Academia Sinica, 11F of AS/NTU Astronomy-Mathematics Building, No.1, Sec. 4, Roosevelt Rd, Taipei 10617, Taiwan}

\author{Jorge K. Barrera-Ballesteros}
\affil{Department of Physics \& Astronomy, Johns Hopkins University, Bloomberg Center, 3400 N. Charles St., Baltimore, MD 21218, USA}
\affil{Instituto de Astronom\'{i}a, Universidad Nacional Aut\'{o}noma de M\'{e}xico, A. P. 70-264, C.P. 04510, M\'{e}xico, D.F., Mexico}

\author{Sebasti\'{a}n F. S\'{a}nchez}
\affil{Instituto de Astronom\'{i}a, Universidad Nacional Aut\'{o}noma de M\'{e}xico, A. P. 70-264, C.P. 04510, M\'{e}xico, D.F., Mexico}

\author{Chin-Hao Hsu}
\affil{Institute of Astronomy and Astrophysics, Academia Sinica, 11F of AS/NTU Astronomy-Mathematics Building, No.1, Sec. 4, Roosevelt Rd, Taipei 10617, Taiwan}
\affil{Department of Physics, National Taiwan University, 10617, Taipei, Taiwan}

\author{Ryan Keenan}
\affil{Institute of Astronomy and Astrophysics, Academia Sinica, 11F of AS/NTU Astronomy-Mathematics Building, No.1, Sec. 4, Roosevelt Rd, Taipei 10617, Taiwan}

\author{Patricia B. Tissera}
\affil{Departamento de Ciencias Fisicas, Universidad Andres Bello, 700 Fernandez Concha, Las Condes, 830000, Santiago, Chile}

\author{M\'{e}d\'{e}ric Boquien}
\affil{Centro de Astronom\'{i}a (CITEVA), Universidad de Antofagasta, Avenida Angamos 601, Antofagasta, Chile}

\author{Y. Sophia Dai}
\affil{Chinese Academy of Sciences South America Center for Astronomy (CASSACA), 20A Datun Road, Beijing 100012, China}

\author{Johan H. Knapen}
\affil{Instituto de Astrofísica de Canarias, 38205 La Laguna, Tenerife, Spain}
\affil{Departamento de Astrof\'{i}sica, Universidad de La Laguna, 38205 La Laguna, Tenerife, Spain}
\affil{Astrophysics Research Institute, Liverpool John Moores University, IC2, Liverpool Science Park, 146 Brownlow Hill, LiverpoolL3 5RF, UK}

\author{Rog\'{e}rio Riffel}
\affil{Departamento de Astronomia, Universidade Federal do Rio Grande do Sul, Campus do Vale, Porto Alegre, RS, Brasil, 91501-970}
\affil{Laborat\'{o}rio
Interinstitucional de e-Astronomia, Rua General Jos\'{e}
Cristino, 77 Vasco da Gama, Rio de Janeiro, Brasil, 20921-400}

\author{Maria Argudo-Fern\'{a}ndez}
\affil{Centro de Astronom\'{i}a (CITEVA), Universidad de Antofagasta, Avenida Angamos 601, Antofagasta, Chile}
\affil{Chinese   Academy   of   Sciences   South   America   Center   for   Astronomy,   China-Chile   Joint   Center   for   Astronomy,Camino El Observatorio 1515 Las Condes, Santiago, Chile}
\affil{Instituto de Fs\'{i}sica, Pontificia Universidad Cats\'{o}lica de Valparas\'{i}so, Casilla 4059, Valparas\'{i}so, Chile}

\author{Ting Xiao}
\affil{Department of Physics, Zhejiang University, Hangzhou 310027, Peopleʼs Republic of China}
\affil{Shanghai Astronomical Observatory, CAS, 80 Nandan Road,Shanghai 200030, China}

\author{Fang-Ting Yuan}
\affil{Shanghai Astronomical Observatory, CAS, 80 Nandan Road,Shanghai 200030, China}

\begin{abstract} % 250 words %
%	250 words.
%Galaxy mergers are the most powerful drivers of galaxy evolution in our local Universe. 
Galaxy interaction is considered a key driver of galaxy evolution and star formation (SF) history.
In this paper, we present  an empirical picture of the radial extent of interaction-triggered SF along the merger sequence.
The samples under study are drawn from the integral field spectroscopy (IFS) survey SDSS-IV MaNGA, including 205 star-forming galaxies in pairs/mergers and $\sim$ 1350 control galaxies.
For each galaxy in pairs, the merger stage is identified according to its morphological signatures: incoming phase, at first pericenter passage, at apocenter, in merging phase, and in final coalescence. 
The effect of interactions  is quantified by  the  global and spatially resolved SF rate relative to the    SF rate of a  control sample selected for each individual galaxy  (\GDSFR{} and \SDSFR{}, respectively).
Analysis of the radial \SDSFR{}  distributions shows that galaxy interactions have no significant impact on  the \SDSFR{} during the incoming phase.  Right after the first pericenter passage,   the  radial \SDSFR{}  profile  decreases  steeply  from enhanced to suppressed activity for increasing galactocentric radius.
Later on,  SF is enhanced on  a  broad  spatial scale out to the maximum radius  we explore ($\sim$ 6.7 kpc) and the enhancement is  in general centrally peaked.
The extended SF enhancement is also observed for systems at their apocenters and in the coalescence phase, suggesting that interaction-triggered SF is not restricted to the central region of a galaxy.
Further  explorations  of a wide range in parameter space of merger configurations (e.g., mass ratio)  are required to constrain the whole picture of interaction-triggered SF.

\end{abstract} % 250 words


\section{Introduction}
%Galaxy interactions are ubiquitous throughout the Universe.
Galaxy interactions significantly alter the star formation  history of galaxies.
It is well established  statistically that the global star formation rate (SFR) increases with decreasing separation between two approaching galaxies  \citep[][]{Lam03,Li08a,Li08b,Scu12,Patt13,Kna15,Pan18}.
The enhanced SFR has been attributed to  the formation of non-axisymmetric structures that torque significant amounts of gas into the central regions, initiating  enhanced (circumnuclear) star formation  \citep[e.g.,][]{Bar91}.

Yet there is  mounting evidence for a  component of extended star formation in interacting galaxies.
The most famous example of such systems is the  Antennae, which  consists of two equal-mass, gas-rich spiral galaxies, NGC 4038 and 4039.
In the Antennae, the majority of the star formation is outside the nuclei \citep{Wan04}.
Other examples include Arp 65 (NGC 90/NGC 93; \citealt{Sen15}),  Arp 299 (IC 694/NGC 3690; \citealt{Alo00}), the IC 2163 and NGC 2207 system \citep{Elm95a,Elm17},  NGC 5291 \citep{Boq07},   IC 1623, NGC 6090, NGC 2623, and the Mice system NGC 4676A/4676B \citep{Wil14,Cor17a,Cor17b,Cor17c}.

Several  observations have attempted to measure the spatial extent of star formation \citep[e.g.,][]{Kna09,Sch13,Wil14,Bar15a,Cor17a,Cor17b,Cor17c,Tho18}.
 \citet{Sch13}  analyzed the spatial extent of star formation in mergers using  60 visually identified galaxy merger candidates  drawn from the 3D-$HST$ survey at $z$ $\sim$ 1.5.
They found that  these systems are  often associated with
the classic circumnuclear  starburst, but their star formation can also be located in  tidal tails. 
\citet{Bar15a}  used  the  information  provided  by  the integral field spectroscopy (IFS) survey   CALIFA (The Calar Alto Legacy Integral Field Area Survey; \citealt{San12})  to carry out the first  statistical study of the impact of the merger event on the star formation distribution in galaxies.
They found moderate enhancement in the global specific star formation rate (sSFR $=$ SFR/$M_{\ast}$, where $M_{\ast}$ is global stellar mass) in the central region of interacting galaxies; however, in the outer regions, the sSFR is similar   to that in the  control sample.
The  extended interaction-triggered star formation is also found in the late stage mergers \citep[e.g.,][]{Boq09,Boq10,Tho18}.  
In addition to those apparently interacting galaxies and final mergers,  \citet{Mcq12} and \citet{Sac18} examined the distribution of star formation in starburst dwarf galaxies.
They found that these galaxies exhibit both extended and concentrated active star formation. 
The widespread star formation might be triggered by external mechanisms, such as interactions/mergers between gas-rich dwarfs or cold gas accretion from the  intergalactic medium  \citep{Nog88,Lel14}.

 Further,   evolution of interaction-triggered star formation distribution has  attracted increasing amounts of attention.
\citet{Ell13}  constrained the  evolution of the extent of interaction-triggered star formation using $\sim$ 11,000 normal interacting galaxies and final mergers, where the merger stage is indicated by the projected separation between two galaxies in a system.
They compared the SDSS fiber (3$\arcsec$) SFR and SFR outside of the fiber (subtracting the fiber SFR from the total SFR derived by the aperture correction). 
Their results showed that the  pre-coalescence  phase  of  the interaction  most  strongly  affects  central  star  formation, while the  final merging process  increases  the  SFR   on  a  broader  spatial scale.
Using  CALIFA and Potsdam Multi-Aperture Spectrometer data for a set of galaxies, \citet{Cor17a,Cor17b,Cor17c} also showed that  the spatial extent and the level of interaction-triggered star formation occur in different time-scales that are connected to the evolutionary stage of the merger.

Despite their relevance, these analyses have several shortcomings.
Previous studies  often contain too few galaxies or only apparently interacting galaxies.
These obstacles  limit our ability to establish a more general picture    of which phase of the merger process   triggers  star formation and where the stars form.
Moreover,  many studies  have been  focused  on high-luminosity major mergers (mass ratio within a factor of 4).
It is not clear  if the result can be generalized to all types of galaxies in pairs and mergers.
Finally, many studies use the projected separation between two galaxies as a merger stage indicator, however,  projected separation is not linearly correlated with merger stages as two galaxies experience several pericenter passages before the final merger, as shown in   simulations \citep[e.g.,][]{Tor12,Mor15,Bus18}.

In this paper, we present an empirical picture of spatially resolved interaction-triggered SFR as a function of merger sequence  using  the IFS  data from the  MaNGA survey (Mapping Nearby Galaxies at Apache Point Observatory; \citealt{Bun15}).
We improve upon previous work by  identifying pairs using both  spectroscopic data and galaxy   morphology. 
This allows us to extend the sample to include  widely separated galaxies in pairs (incoming systems or systems at their apocenter).
The unprecedented number of galaxies with MaNGA  data  allows for  a carefully selected control sample for each individual galaxy and   a quantification of interaction-triggered star formation.
In addition, we design a scheme for classifying the stage of an interaction based on the morphological appearance of the system and thus, do not rely on the nuclear separation.

This paper is organized as follows.
In Section \ref{sec_merger_stage}, we introduce the scheme for merger stage classification.
Data and analysis are presented  in Section \ref{sec_data}.
In Section \ref{sec_result}, we show the dependence of both the integrated and spatially-resolved star formation rate on the merger sequence.
The results are discussed in Section \ref{sec_discussion} and summarized in Section \ref{sec_summary}.
Throughout this paper, we assume $\Omega_\mathrm{m}$ $=$ 0.3, $\Omega_\mathrm{\Lambda}$ $=$ 0.7, $H_{0}$ $=$ 70 km s$^{-1}$ Mpc$^{-1}$.

\section{Merger Sequence}
\label{sec_merger_stage}
As mentioned earlier, the nuclear separation between two galaxies does not linearly correlate with merger stages, therefore, we determine the merger stages via visual inspections of the  $gri$ composite images observed by the  2.5-m Telescope of the Sloan Digital Sky Survey (SDSS; \citealt{Gun06}).
Interactions between galaxies are classified according to the following scheme:

\begin{itemize}
	
	\item {\bf Stage 1} -- well-separated pair  which do not show any morphology distortion (i.e., incoming pairs, before the first pericenter passage),
		
	\item {\bf Stage 2} -- close pairs showing strong signs of interaction, such as tails and bridges (i.e., at the first pericenter passage),

	\item {\bf Stage 3} -- well-separated pairs, but showing weak morphology distortion (i.e.,  approaching the apocenter or just passing the apocenter),

	\item {\bf Stage 4} --  two components strongly overlapping with each other and show strong morphological distortion (i.e., final coalescence phase), or single galaxies with obvious tidal features such as tails and shells (post-mergers),
		
\end{itemize}

\begin{figure*}%
	\centering
	\includegraphics[width=0.7\textwidth]{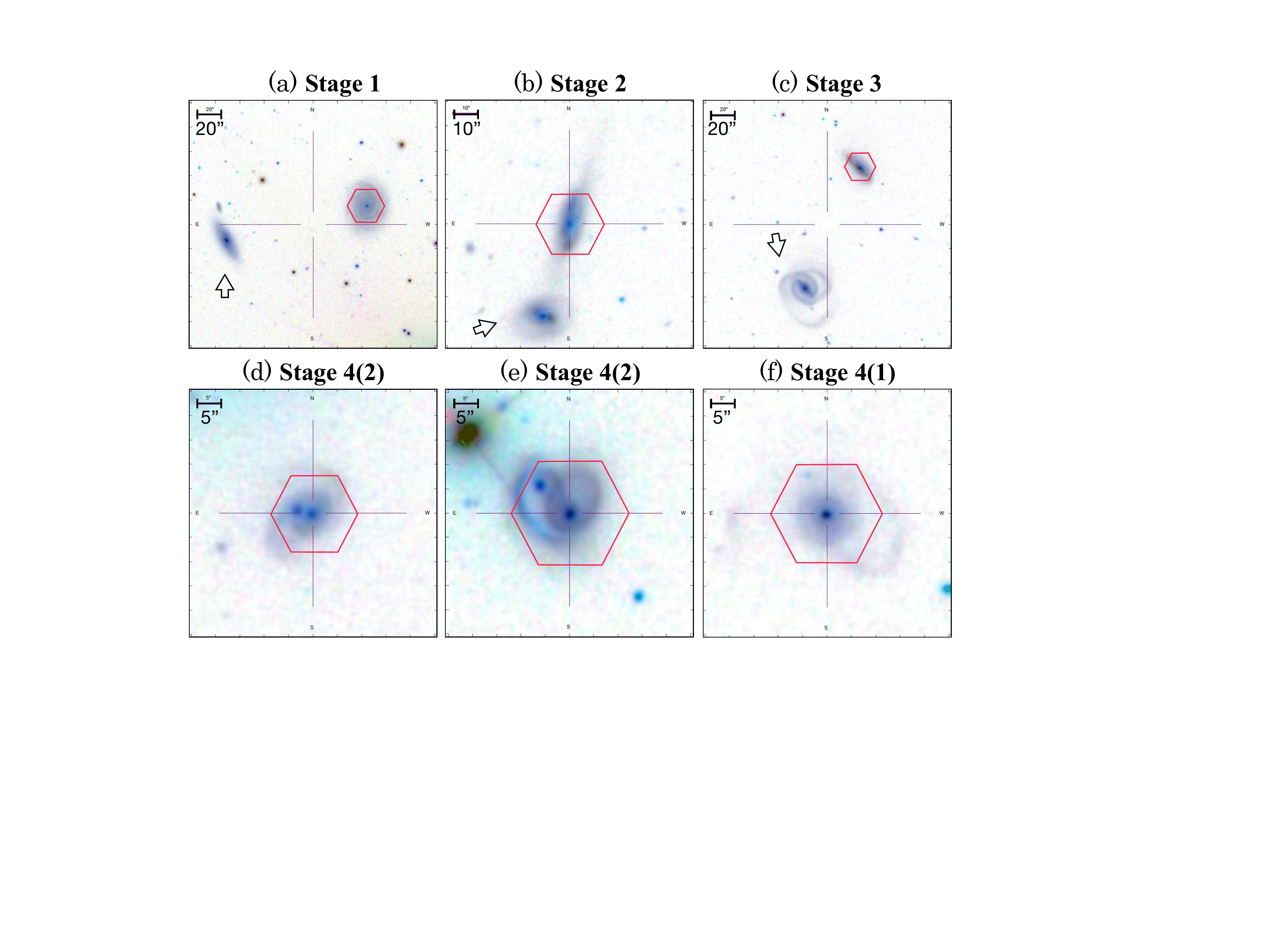}
	\caption{Examples of galaxies in pairs and mergers at different merger stages (\S\ref{sec_merger_stage}).  In the upper panels, the red hexagons show the coverage of the MaNGA IFU. The companions are indicated by arrows. Panel (a)--(c) preset the examples of Stage 1 to 3, respectively. The lower panels present the diverse morphology of galaxies at the (post-)coalescence phase (Stage 4). Panel (d) and (e) are Stage 4(2) (double nuclei), panel (f) is Stage 4(1) (single nucleus). In the main text,  we refer the stages as \textit{S1} --  \textit{S3}, and  \textit{S4(2)} and  \textit{S4(1)}, respectively.  The linear scale of the image is indicated in the upper-left corner of each panel.}%
	\label{fig_stages}%
\end{figure*}

Examples of each stage are presented in Figure \ref{fig_stages}. 
The scheme is analogous to the Toomre Sequence (\citealt{Too77}; see also \citealt{Vel02} and \citealt{Bar15b}) and the morphological evolution of simulated mergers (e.g., Figure 8 in \citealt{Tor12} and Figure 2 in \citealt{Mor15}).
Although this stage classification may not reflect the full merging process, it is useful for a demonstration purpose (more discussion, see Section \ref{sec_caveat_stage}). 

\section{Data and Analysis}
\label{sec_data}

\subsection{MaNGA}
MaNGA  is  part  of  the  fourth  generation  Sloan  Digital Sky Survey \citep[SDSS-IV;][]{Bla17}, and aims to  survey for $\sim$ 10,000 galaxies with a median redshift ($z$) of 0.03 by 2020.
The observations are carried out with  integral field units (IFUs)  of different sizes, varying in diameter from 12$\arcsec$ (19 fibers) to 32$\arcsec$ (127 fibers). 
%The IFUs are installed in six SDSS cartridges. 
%Each MaNGA cartridge has 17 science IFUs and 12 seven-fiber IFUs for calibration. 
The IFU sizes and the number density of galaxies on the sky were designed jointly to allow more efficient use of IFUs (e.g., to minimize the number of IFUs that are unused due to a tile with too few galaxies), and to allow us to observe galaxies in the $z$ range to at least 1.5 effective radii ($R_{e}$), where $R_{e}$ is the radius containing 50\% of the light of the galaxy measured at $r$-band.
The  fibers are fed into the dual beam BOSS spectrographs \citep{Sme13}, covering the wavelength range from 3600\AA\, to 10300 \AA.
The spectral resolutions vary from  R $\sim$ 1400 at 4000\,\AA\, to R$\sim$ 2600 around 9000\,\AA\, \citep{Dro15,Yan16,Yan16b}.
The point spread function (PSF)  is $\sim$ 2.5$\arcsec$, corresponding to 1.8 kpc at the median redshift of the current MaNGA sample (0.036).
For more detail on the MaNGA setup,  we refer the reader to \citet{Dro15} for the IFU fiber feed system, to \citet{Wak17}  for the sample selection, to \citet{Law15} for the observing strategy, and  to \citet{Law16} and \citet{Wes19} for the   MaNGA data reduction and data analysis pipelines, respectively.
We select galaxies from a sample of 4691 galaxies observed by MaNGA within the first $\sim$ 4 years of operation, corresponding to the SDSS data release 15 \citep{Agu18}.

\subsection{Local $M_{\ast}$ and SFR   Measurements}
\label{sec_pipe3d}
The reduced   MaNGA datacubes are analyzed using the Pipe3D pipeline  to extract the  physical parameters from each of the spaxels of each galaxy \citep{San16a,San16b,San18}.
Each spaxel has an area of 0.5$\arcsec$ $\times$  0.5$\arcsec$.
Pipe3D fits the continuum with stellar population models and measures the nebular emission lines.
Here we briefly describe the procedures. 

The stellar continuum is  modeled using a simple-stellar-population (SSP) library  with 156 SSPs, comprising 39 ages and 4 metallicities \citep{Cid13}. 
Before the fitting,  spatial binning is  performed to reach a signal-to-noise ratio (S/N) goal of 50 across the field of view. 
Then  the stellar population fitting was applied to the coadded spectra within each spatial bin.
Finally, the stellar population model for spaxels with continuum S/N $>$ 3 is derived by re-scaling the best fitted model within each spatial bin to the continuum flux intensity  in  the  corresponding spaxel. 
The stellar mass  is obtained using the stellar populations derived for each spaxel, then normalized to the physical area of a spaxel to get the surface density ($\Sigma_{\ast}$) in  M$_{\sun}$ kpc$^{-2}$.

Then the stellar-population models  are  subtracted from the data cube to create an emission line cube. 
%The emission line fluxes were measured spaxel by spaxel.
SFR  is derived using   H$\alpha$.  
Since H$\alpha$ may be powered by various sources (e.g., star formation,  evolved stars, and AGNs),  we use   excitation diagnostic diagrams \citep{Bal81,Kew01,Kau03,Cid13}   and an H$\alpha$ equivalent width cut of $>$ 6 \AA\ \citep{San14} to pick up star-forming regions (for more discussions on the equivalent width of H$\alpha$ and  the nature of the line emitting gas, see \citealt{Lac18}).
To this end, we  limit the  analysis to spaxels with S/N $>$ 3 for H$\alpha$, H$\beta$, [\ion{O}{3}] and [\ion{N}{2}]. 
Only star-forming spaxels are used for the analysis of this work.
We will restricted our objects to star-forming galaxies (\S\ref{sec_global_sm_sfr}),  whose H$\alpha$ emission are dominated by star-forming regions \citep{Pan18b}.
Therefore $\sim$ 93\% of spaxels  in our sample galaxies are star-forming spaxels.
The method described in \citet{Vog13} is used to compute the reddening using the Balmer decrement at each spaxel.
The extinction-corrected H$\alpha$ luminosity is converted into SFR surface density  ($\Sigma_\mathrm{SFR}$ in M$_{\sun}$ yr$^{-1}$ kpc$^{-2}$) using the  calibration from \citet{Ken98}. 
Inclination correction  is  applied to   $\Sigma_{\ast}$ and $\Sigma_\mathrm{SFR}$ of all spaxels of a galaxy equally.

\subsection{Global $M_{\ast}$ and SFR   Measurements}
\label{sec_global_sm_sfr}
The global stellar mass ($M_{\ast}$) is taken from the MPA/JHU catalog\footnote{https://wwwmpa.mpa-garching.mpg.de/SDSS/DR7/}, where
 $M_{\ast}$ is estimated by fitting stellar population models from \citet{Bru03} to the ugriz SDSS photometry, following the method of \citet{Kau03} (see \citet{Gar19} for more discussions on the radial structure of the mass-to-light ratio). 
The $M_{\ast}$ have been found to agree with other estimates \citep[e.g.,][]{Tay11,Men14,Cha15}. 
To be consistent with previous studies of SDSS pairs \cite[e.g.,][]{Scu12,Patt13,Ell13}, this work focuses on the galaxies in systems with $M_{\ast}$ ratio less than 10 (from 1:10 to 10:1). 
%For this reason, we cross match the companions and MPA/JHU catalog to pick up the systems whose mass ratios are computable.
%Applying the mass ratio selection, the number of galaxies in pairs become 192.
In principle, the total $M_{\ast}$  can also be computed for the MaNGA sample by integrating across all the spaxels in the IFU \citep[e.g.,][]{Can16,San18}. 
However, as more than 90\% of the companions are not observed in MaNGA,   we cannot estimate the $M_{\ast}$ ratio from the MaNGA data alone.
For this reason, the $M_{\ast}$ from the MPA/JHU catalog is adopted.
The MPA/JHU catalog assumes a Kroupa IMF \citep{Kro01}, while Pipe3D adopts a Salpeter  IMF.
So in order to convert $M_{\ast}$ from a Kroupa IMF to a Salpeter IMF, 0.2 dex have to be added to the $M_{\ast}$, which corresponds to a factor of 1.6.
After applying the conversion factor,  the mean difference between the MPA/JHU and integrated MaNGA $M_{\ast}$ is 0.015 dex.

To fairly compare the results of global and local SFR, the   star formation per spaxel from MaNGA are  coadded to derive the global  SFR of the galaxies  \citep{San18}.
The MPA/JHU catalog also provides the global SFR measurements. 
The mean difference between the MPA/JHU SFR (after IMF conversion) and integrated Pipe3D SFR is somewhat larger, 0.17 dex, consistent with comparisons in other papers \citep[e.g.,][]{Ell18,Spi18}.
The significant difference in the two SFRs is most likely due to the use of the aperture correction to the 3$\arcsec$ fibers in SDSS missing star formation which is present in the MaNGA IFUs.
To avoid any uncertainties associated with aperture correction \citep[e.g.,][]{Ric16,Spi18} and to be consistent with the value of local SFR, we use the integrated MaNGA SFR in this work.

Galaxies that are quenching or quenched are removed from this work. A criterion of log(sSFR/yr$^{-1}$) $>$ -11 is applied to select star-forming galaxies, where the specific SFR (sSFR) is defined as SFR/$M_{\ast}$.
Varying the criteria between log(sSFR/yr$^{-1}$) $=$ -11.0 and -10.5 will not change our main conclusions, but the number of galaxies in pairs would be reduced by $\sim$ 20\%. 
 We emphasize that we do not require our sample galaxies to have star-forming companions.
We include star-forming galaxies interacting with an early-type galaxy, for instance.
The dependence of interaction-triggered star formation on mass ratio and properties of companion will be discussed in a separate paper. %} 

 Finally, it is worth noting  that the star formation rate in this work is calculated by H$\alpha$ luminosity, tracing the ongoing star formation ($<$ 100 Myr), while the extension and enhancement of star formation can occur in different time-scales. 
We refer the reader to \citet{Pan15} and \citet{Dav15} for the comparisons of interaction-triggered SFRs using observational tracers which probe different star formation time-scales and \citet{Cor17a,Cor17b,Cor17c} for the constraints onto interaction-triggered star formation history using population synthesis.

\subsection{Identifying Galaxies in Pairs and Mergers in MaNGA}
\label{sec_pair_select_all}
\subsubsection{Projected Separation and Velocity Difference}
\label{sec_pair_select}
The galaxies in pairs or mergers (\emph{p/m}) are defined as  galaxies with a spectroscopic companion\footnote{The field of view of MaNGA observations can only cover one galaxy in an interacting system except for some  very close pairs of galaxies and galaxies in the coalescence phase. Therefore we refer to the samples as galaxies in pairs or mergers.}.
The parent sample is made up of 641,409   nearby galaxies  from the NSA,  primarily based on the SDSS DR7 main galaxy sample \citep[][]{Aba09} , but incorporating data from additional sources. 
NSA is also the parent catalog  for target selection for MaNGA \citep{Wak17}.

Galaxies in pairs must have a companion at a projected separation  $<$ 50 kpc h$^{-1}$ (or 71.4 kpc)   and  a line-of-sight velocity difference  $<$ 500 km s$^{-1}$ \citep[e.g.,][]{Pat02,Lin04}. 
This results in a  sample of 34,478 galaxies in pairs or multiples. 
If several companions are found for a given  galaxy,  the companion with the smallest separation is selected. 
Then we  cross match the NSA galaxies in pairs  and the  4,691 MaNGA galaxies, yielding a sample of 682 galaxies in pairs. We note that in more than 90\% of the cases, only one of the two components of a galaxy pair is observed by MaNGA.
Applying the mass ratio and sSFR selections (Section \ref{sec_global_sm_sfr}),  the number of galaxies in pairs become 109, including 38 in Stage 1, 10 in Stage 2, 58 in Stage 3, and 3 in Stage 4.

\subsubsection{Morphology}
\label{sec_stage_morph}
One caveat to the identification of galaxies in \emph{p/m} by the projected separation and  velocity difference is that if the two  components are too close (i.e., Stage 2 and Stage 4) to be de-blended by SDSS or do not have two separate spectroscopic redshifts, they will be  missed in the spectroscopic determination.
Another caveat  is that  post-mergers (i.e., single galaxies) are missed by the selection criteria. 

We  visually inspect the SDSS \emph{gri} composite images of all MaNGA galaxies and  recover these closest interaction systems.
The number of Stage 2 and Stage 4 galaxies then increases to 24 and 85, respectively.
It should be noted that the galaxies in Stage 2  may span a narrower mass ratio range compared to other stages because in principle similar-$M_{\ast}$ interaction can generate stronger signs of tidal features.
However, it is impossible to obtain the mass ratio of galaxies at Stage 4 without modeling the properties of progenitors.
Besides, one has to keep in mind that a larger diversity of morphologies is observed for galaxies at Stage 4 compared to other stages, such as double nuclei with  tidal features (Figure \ref{fig_stages}(d) and (e)) and single nuclei with   shell structures  (\ref{fig_stages}(f)).
Of the 85 galaxies in Stage 4, 13 and 72 show double nuclei and single nucleus, respectively.

\subsubsection{Summary of the Sample of Galaxies in P/M}
The sample   for this paper comprises  205 galaxies in \emph{p/m}. 
Of these,   53\%  are selected based on the   spectroscopic data (Section \ref{sec_pair_select}) and 47\% based on their morphology (Section \ref{sec_stage_morph}).
For the galaxies with spectroscopic companions, $\sim$ 57\% are primary galaxies (the higher $M_{\ast}$ one in a pair), while $\sim$ 43\% are secondary galaxies.
The fraction (and number) of galaxies in Stage 1 -- Stage 4 are 19\%(38), 12\%(24), 28\%(58), and 41\% (85), respectively.
Of the galaxies in Stage 4, 15\%(13) of them have visually  double nuclei, while 85\%(72) have one. 
Hereafter, we refer to  Stage 1 -- 4 as  \textit{S1} -- \textit{S4}, respectively.
For \textit{S4}, we use \textit{S4(2)} and \textit{S4(1)} to represent the sub-categories for double nuclei and single nucleus, respectively.

The small numbers of \textit{S2} and \textit{S4(2)} in our sample are presumably due to the short time-scale during which the galaxies exhibit the features to be identified as  \textit{S2} and \textit{S4(2)}.
The two nuclei are separated by a few to several kpc  in these galaxies.
According to  simulations, the duration of such phases are several to tens of times shorter than other stages \citep[e.g.,][]{Tor12,Ji14}.

%The sSFR cut yields a sample of 117 star-forming galaxies in pairs.

\subsection{Control Sample of Isolated Galaxies}
\label{sec_control}
We define a control sample to quantify the effect of interactions.
To construct a reliable control sample, in addition to the spectroscopic determination, we  also   use the  ``P-merger'' parameter from  Galaxy Zoo to remove  potential interacting galaxies through their morphology \citep{Dar10a,Dar10b}.
P-merger quantifies the probability  that an object is a merger.
The value ranges from 0, an object looks nothing like a merger, to 1, an object  looks  unmistakably so.
The control galaxies should have no spectroscopic companion,  \emph{and} P-merger $=$ 0,  \emph{and} log(sSFR/yr$^{-1}$) $>$ 11.
A total of 1348 MaNGA star-forming galaxies are selected as control galaxies.
This pool of control galaxies will be used to select the control galaxies for a given galaxy (\S\ref{sec_offset}).

\begin{figure}%
	
	\includegraphics[width=0.45\textwidth]{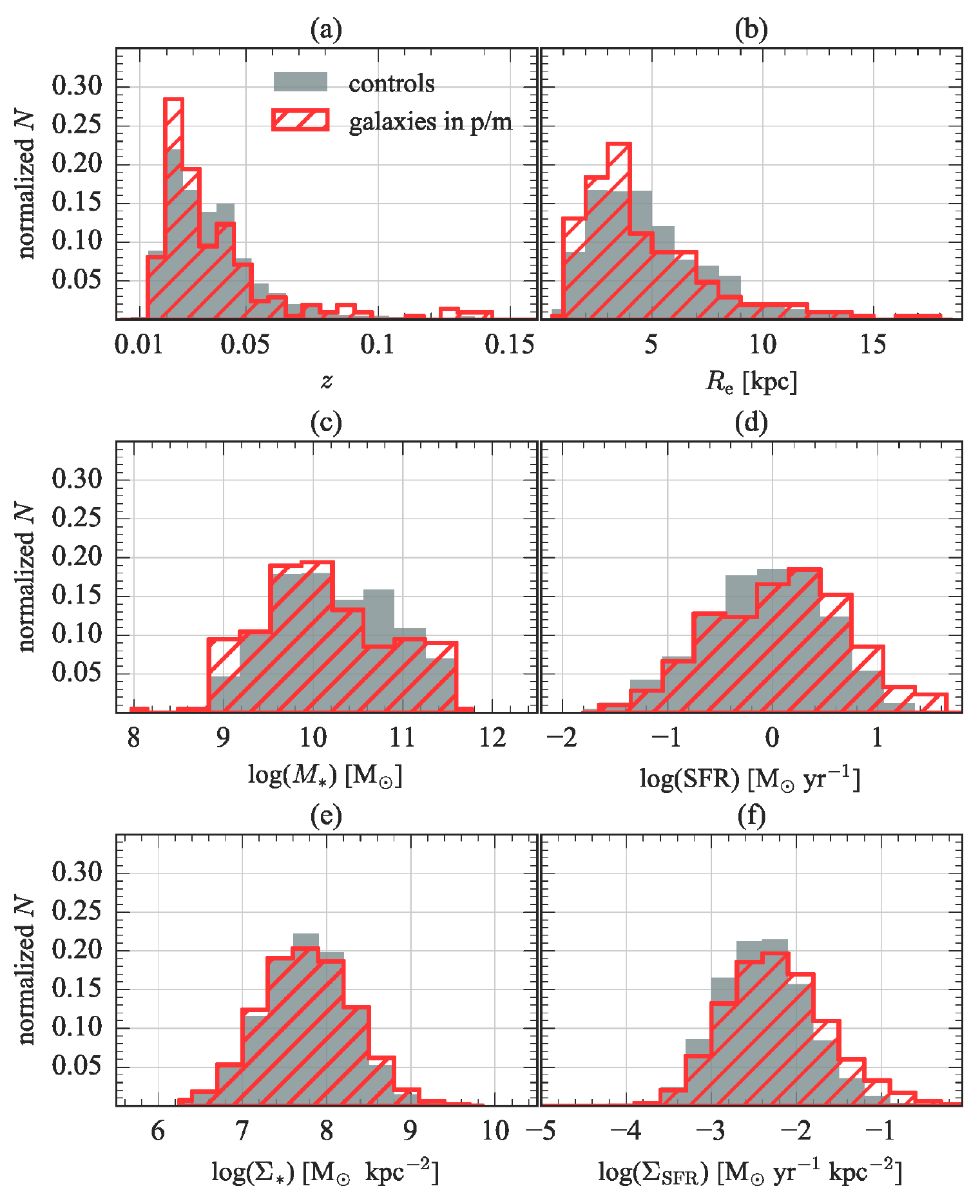}
	\caption{Distributions of galaxy global and spatially-resolved properties: (a) spectroscopic redshift ($z$),  (b) effective radius  measured at $r$-band ($R_\mathrm{e}$), (c) global stellar mass ($M_{\ast}$) taken from the MPA/JHU catalog (\S\ref{sec_global_sm_sfr}),  (d) global star formation rate (SFR)  taken from MaNGA (\S\ref{sec_global_sm_sfr}), (e) spaxel-wide stellar mass surface density ($\Sigma_{\ast}$; \S\ref{sec_pipe3d}), and (f)  spaxel-wide star formation rate surface density ($\Sigma_\mathrm{SFR}$; \S\ref{sec_pipe3d}). Filled and hatched histograms indicate the galaxies in the control pool and galaxies in pairs and mergers, respectively.}%
	\label{fig_Gal_prop}%
\end{figure}

The $z$ and $R_{e}$ distributions (from NSA) of galaxies in \emph{p/m} and controls are shown in Figure \ref{fig_Gal_prop}(a) and (b), along with $M_{\ast}$, SFR, $\Sigma_{\ast}$, and $\Sigma_\mathrm{H_{2}}$ in panel (c) -- (f), respectively. 
The galaxies in \emph{p/m} and controls are represented by the hatched and solid histograms, respectively.
The distributions show good overall similarity between the  galaxies in \emph{p/m} and control galaxies.

\subsection{Quantification of Interaction-triggered Star Formation}
\label{sec_offset}
In order to fairly compare the properties of the galaxies in \emph{p/m} and controls,  we compute ``offset''  quantities.
Our approach follows closely that of \citet{Ell18} for the analysis of the spatially-resolved main sequence.
The offset values are computed for both control galaxies and galaxies in \emph{p/m}.
The distribution of the offset values of  controls is useful to give an idea of the intrinsic scatter of the offset values.

\subsubsection{Global SFR Offset: \GDSFR{}}
\label{sec_global_offset}
Each  galaxy  is matched in $z$, $M_{\ast}$, and $R_{e}$ with a minimum of three control galaxies  from the pool of controls (\S\ref{sec_control}).
We allow shared control galaxies for different galaxies.
The initial tolerance of $z$, $M_{\ast}$, and $R_{e}$ are 0.005, 0.1 dex and 20\%, respectively.
The criteria are allowed to grow by 0.005, 0.1 dex, and 5\%, respectively, until the minimum required number of control galaxies  is reached. 
In practice, $\sim$ 90\% of galaxies are successfully matched to at least three controls without the need to grow the tolerances. 
Around 9\% of galaxies require only one growth  step in order to reach the requirement of three matched controls.
The remaining galaxies can find sufficient controls in the third or fourth growth  steps.
In general, the number of matched controls  exceeds the minimum requirement of three, with an average of  10 matches per galaxy.
The ``offset'' of global SFR in  logarithm scale (\GDSFR{})  is defined as,
\begin{equation}
\Delta\mathit{log}{\mathrm{SFR}}=\mathit{log}\mathrm{SFR}-\mathit{log}\,\mathrm{median(SFR_{controls})},
\end{equation}
where $log\mathrm{SFR}$ is the SFR of the galaxy in question and $log$ median(SFR$_\mathrm{controls}$) is the median SFR of its  control galaxies in  logarithm scale.
The \GDSFR{} is calculated for both the galaxies in \emph{p/m} and for all controls.
We should emphasize that as this is taken in the logarithm form, it really is a ratio of a value of the galaxy in question against the median value of its controls.
A positive offset  represents an enhancement of  global SFR with respect to the controls, and vice versa.

\subsubsection{Local sSFR Offset: Radial \SDSFR{} Distribution}
\label{sec_local_offset}
For each galaxy, we calculate the ``offset'' of the radial sSFR    distribution with respect to  its controls selected in Section \ref{sec_global_offset} (i.e., according to the global galaxy properties)  using the derived $\Sigma_{\ast}$ and $\Sigma_\mathrm{SFR}$.
For each galaxy, we first calculate its own de-projected radial $log$sSFR($r$)  distribution  with a radial bin of 0.15$R_{e}$.
Then the  radial \SDSFR{}  distribution  of the galaxy in question is computed by subtracting the median radial  $log$sSFR($r$)  distribution  of its controls from its own $log$sSFR($r$)  distribution. Specifically,  at each radial bin $r$:

\begin{equation}
\Delta\mathit{log}\mathrm{sSFR}(r)= \mathit{log}\mathrm{sSFR}(r)-\mathit{log}\,\mathrm{median}(\mathrm{sSFR}(r)_\mathrm{controls}),
\end{equation}
  where $log\mathrm{sSFR}(r)$ represents the logarithm of local sSFR of  the galaxy in question and $\mathit{log}\,\mathrm{median}(\mathrm{sSFR}(r)_\mathrm{controls})$ is the median local sSFR of control galaxies in  logarithm form.
We express the galactocentric distance in units of $R_{e}$ to allow to produce median profiles from the control galaxies.
The radial  distribution  are  computed out to 1.5 $R_{e}$ because the radial coverage of MaNGA fiber bundles is $\geq$ 1.5 $R_{e}$.
As for the global \GDSFR{}, the radial \SDSFR{} distribution  is calculated for both galaxies in \emph{p/m} and controls.

\section{Results}
\label{sec_result}

\subsection{Global SFR Properties}
\label{sec_global}

To compare with the literature, we first show the results of global star formation properties in our sample. 
Figure \ref{fig_Global_dSFR_hist_full} compares the \GDSFR{} distribution of the whole sample of galaxies  in \emph{p/m} (hatched histogram) and  the galaxies in the control pool (solid histogram).
The distribution of the control sample peaks at zero (with a median value of -0.007 dex), confirming that  the approach we have taken to calculate the \GDSFR{} is valid. 
The width  of the distribution of the controls indicates the intrinsic spread of \GDSFR{};  the  standard deviation of the  \GDSFR{} values is 0.34 dex.

The distribution of the controls is symmetric, whereas the distribution of galaxies in \emph{p/m}  is skewed towards larger values of \GDSFR{}.
For the whole data set of galaxies in \emph{p/m} we obtain a  \GDSFR{} standard deviation of   0.44 dex.
While the distributions indicate  higher \GDSFR{} values for galaxies in \emph{p/m} than for those in the control pool,  statistically, the SFR only increases by a limited factor.
The  median \GDSFR{} for galaxies in \emph{p/m} is  0.21 $\pm$ 0.03 dex (around $\times$ 1.6),  where the error bar is the standard error of the mean.
Various observational studies also find a rather limited increase in the SFR during galaxy interactions \citep[e.g.,][]{Lin07,Kna09,Scu12,Ell13,Kna15,Pan18}.
The factor of $\sim$ 1.6 increment in the SFR is the average  value for  galaxies at any instant of the merging process, so it likely indicates that peak SFR values are higher than this. 
In the next section, we examine the \GDSFR{} values  along the merger sequence.

\subsubsection{\GDSFR{} versus Merger Stage}
\label{sec_global_stage}
\begin{figure}
	\begin{center}
		\subfigure[]{\label{fig_Global_dSFR_hist_full}\includegraphics[scale=0.6]{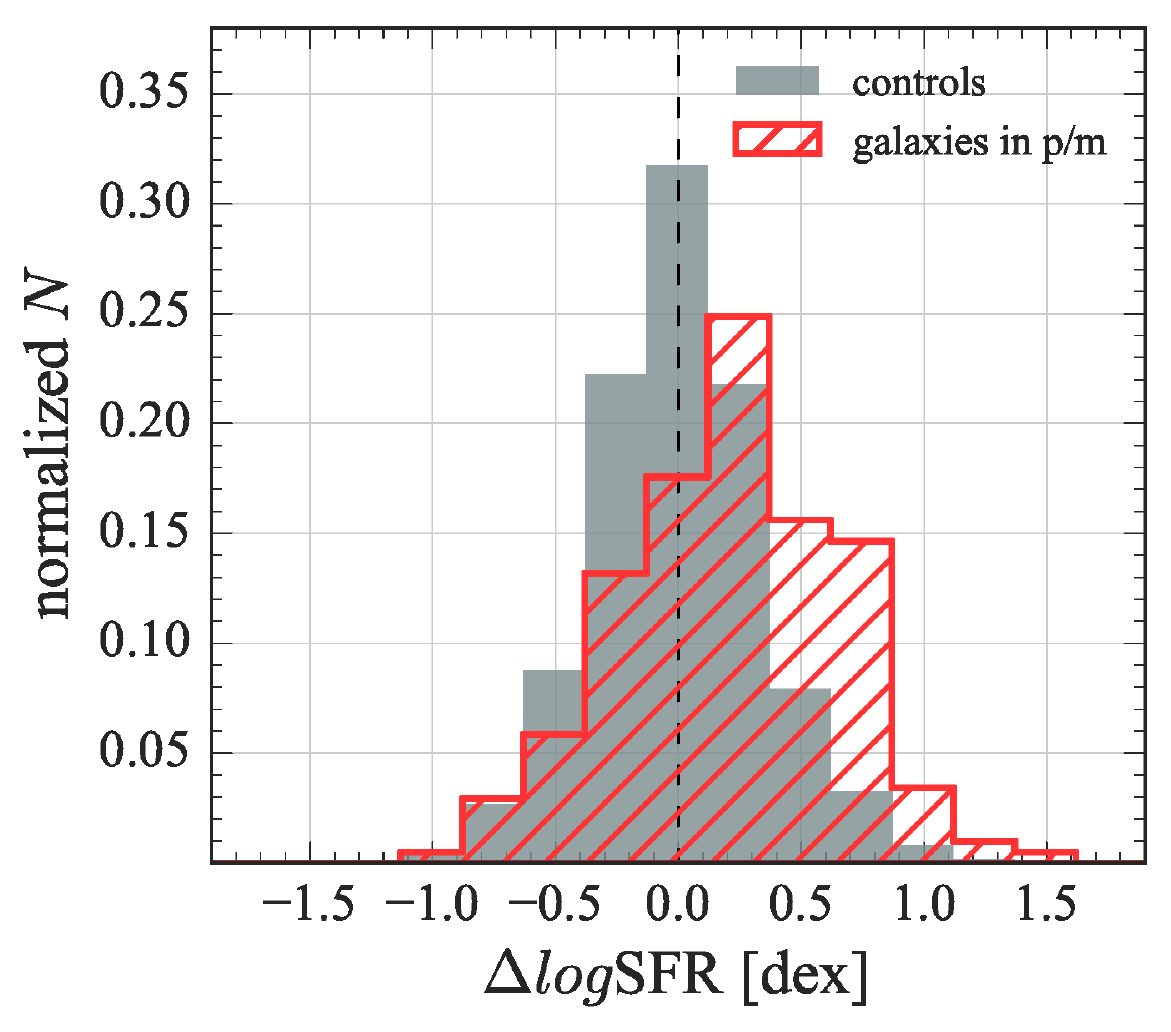}}
		%\subfigure[]{\label{fig_Global_dSFR_hist}\includegraphics[scale=0.5]{Global_dSFR_hist.pdf}}
		\subfigure[]{\label{fig_all_g}\includegraphics[scale=0.6]{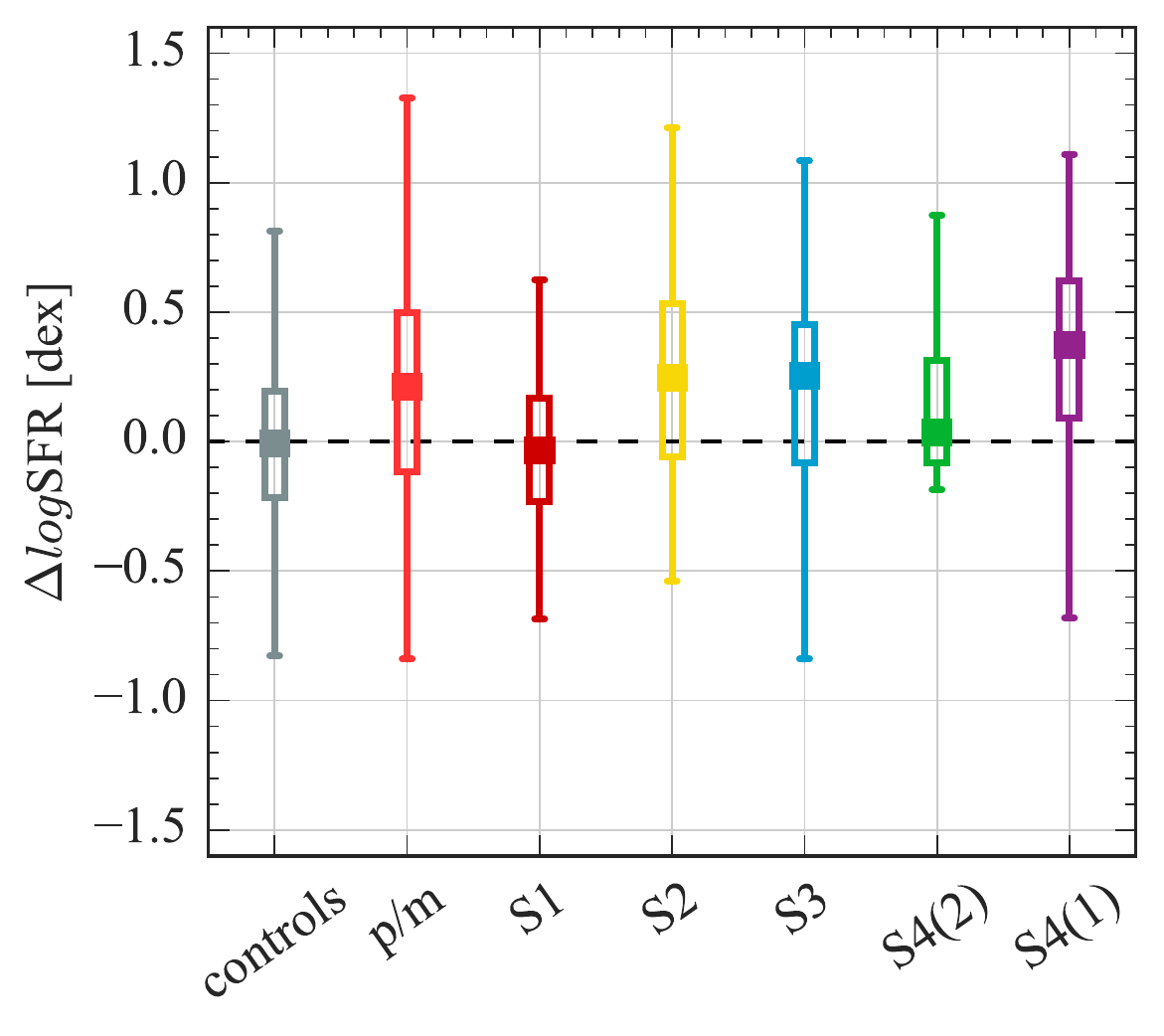}} 
	\end{center}
	\caption{
		(a) Distributions of \GDSFR{} for galaxies in \emph{p/m} (hatched histogram) and controls (solid histogram). The vertical dashed line indicate zero enhancement. (b) Boxplots showing distribution of \GDSFR{} values for different categories of galaxies.	In each boxplot, the median is indicated by the solid squares in the middle. The ends of the box are the upper and lower quartiles (the interquartile range, IQR), 50\% of the sample is located inside  the box. The two whiskers (vertical lines) outside the box extend to 1.5 $\times$ IQR.
		Boxplots from left to right represent the distribution for the galaxies in the control pool, all galaxies in \emph{p/m}, and \textit{S1} -- \textit{4(2)}, respectively.	}
	\label{fig_global_all}
\end{figure}

The  \GDSFR{} values as a function of merger stage is illustrated in Figure \ref{fig_all_g}, where the whole sample of control  galaxies in the \emph{p/m} sample is plotted as well for reference.
The boxplots represent the distribution of the \GDSFR{} values for different categories of galaxies.
In each boxplot, the median is indicated by the solid squares in the middle. The ends of the box are the upper and lower quartiles (the interquartile range, IQR), 50\% of the sample is located inside  the box. The two whiskers (vertical lines) outside the box extend to 1.5 $\times$ IQR.
Boxplots from left to right represent the distribution for the galaxies in the control pool, all the galaxies in \emph{p/m}, and \textit{S1} -- \textit{4(2)}, respectively.

Although the individual galaxies span a wide range of  \GDSFR{}, the median \GDSFR{}s for each of the different stages are still packed in a relatively narrow range of values.
The median \GDSFR{} are -0.04 $\pm$ 0.06 dex for \textit{S1} ($\times$ 0.91), 0.24 $\pm$ 0.10 dex for \textit{S2} ($\times$ 1.73), 0.25 $\pm$ 0.07 dex for \textit{S3} ($\times$ 1.77),  0.04 $\pm$ 0.09 dex for \textit{S4(2)} ($\times$ 1.09), and 0.37 $\pm$ 0.05 dex for \textit{S4(1)} ($\times$ 2.34).
During the incoming phase, the reported SFRs are statistically indistinguishable from those of the control sample.
Then the  median SFR increases since the first passage (\textit{S2} -- \textit{S3}).
The median SFR in the merging phase (\textit{S4(2)}) is not significantly enhanced, which is somewhat surprising.
It could be the result of supernova feedback triggered during the enhanced SFR of the previous phases that quench the star formation activity during this period.
But, due to low number statistics and the large spread in the values, nothing conclusive can be said in this respect  for this specific stage. 
Finally, the median SFR peaks at the  coalescence phase, although it is only by a factor of $\sim$ 2.3.
 Our results are comparable to those reported by \citet{Kna15}, who also quantify the global SFR variation  as a function of  morphologically defined interaction class  using the data from the S$^{4}$G survey (see their Figure 4).

\subsubsection{\GDSFR{} versus the Projected Separation}
\label{sec_global_sep}
The projected separation ($r_{p}$) between two galaxies in a pair is the most accessible and widely used indicator of the stage of an interaction. 
This indicator has been used to understand the role of galaxy interaction in triggering star formation \citep{Lam03,Li08a,Li08b,Scu12,Ell13,Pat13,Dav15,Pan18} and AGNs \citep{Alo07,Li08b,Kos12,Ell13,Gor17}  as well as in altering the cold gas properties \citep{Com94,Gao99,Scu15,Ell18,Pan18}, metallicity \citep{Mic08,Scu12,Guo16},  and morphology  \citep{Cas13} of galaxies.
In this section, we revisit the relation between \GDSFR{} and $r_{p}$ by considering the merger stages.

\begin{figure*}
	\begin{center}
		\includegraphics[scale=0.55]{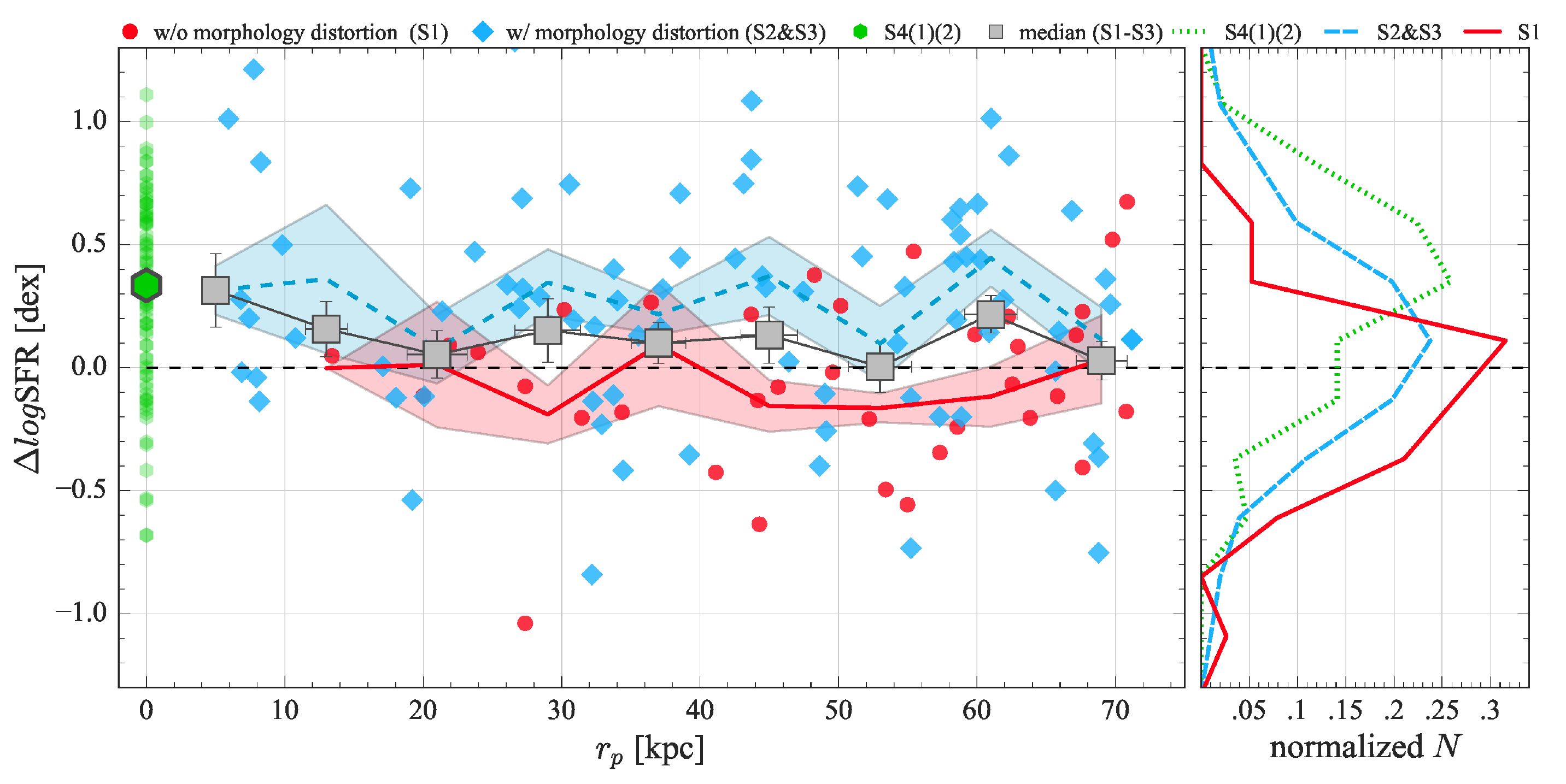}
	\end{center}
	\caption{Left: Offset global SFR (\GDSFR{}) for galaxies in \emph{p/m}  as a function of projected separation ($r_{p}$). Red circles represent galaxies in pairs which show no morphological distortion (i.e., Stage 1). Blue diamond are galaxies in pairs which show morphological distortion (Stage 2 and 3). The galaxies in Stage 4 are shown as green pentagons, and are placed at $r_\mathrm{p}$ $=$ 0. The blue dashed and red solid   lines represent the median \GDSFR{} of galaxies with and without morphological distortion per $r_{p}$, respectively. The shaded areas correspond to the standard error of the mean.  The large gray squares and green hexagon indicate the median \GDSFR{} of galaxies in pairs (Stage 1 -- Stage 3) and in the coalescence phase (Stage 4), respectively. Error bars are standard error of the mean. The horizontal dashed line indicates the zero line.  Right: Normalized \GDSFR{} distributions for galaxies in different merger stages. The solid, long dashed, and short dashed line represent galaxies in Stage 1, Stage 2\&3, and Stage 4, respectively. 
		}
	\label{fig_Projsep_large}	

\end{figure*}

The calculated \GDSFR{} values  are plotted against  $r_{p}$ in the left panel of Figure \ref{fig_Projsep_large}.
In this plot, galaxies in pairs are classified as those with (blue diamonds) and without (red circle) morphological  distortions; the former category consists of galaxies at \textit{S2} and \textit{S3}, while the latter are \textit{S1} objects.
Galaxies in the coalescence phase (\textit{S4}) are plotted at   $r_{p}$ $=$ 0 kpc. 
The blue dashed and red solid   lines represent the median \GDSFR{} of galaxies with and without morphological  distortion per $r_{p}$ bin, respectively.
The large squares and hexagon   indicate the median \GDSFR{} per $r_{p}$ of all of the galaxies  in pairs  (\textit{S1 -- S3}) and in the coalescence phase (\textit{S4}), respectively.

The most important result from Figure \ref{fig_Projsep_large} is that  the galaxies with morphological  distortion (\textit{S2 -- S4}) tend to be found in the positive regime.
This can evidently be  seen in the  histogram of \GDSFR{}  in the right panel (blue dashed and green dotted histograms).
Moreover,  both galaxies in pairs with (\textit{S2} and \textit{S3})  and without (\textit{S1}) morphological  distortion clearly span a wide range in $r_{p}$.
As a result, the  median \GDSFR{} of galaxies with morphological  distortion (blue line) lies above the line of galaxies without  morphological  distortion  (blue line) at almost all $r_{p}$.
Combining these two distinct trends yields a relation very similar (in terms of both profile and values of  \GDSFR{}) to the well-known \GDSFR{} versus $r_{p}$ relation seen in   larger samples \citep[e.g.,][]{Scu12,Ell13}, with an increase in  \GDSFR{}  at the smallest separations.
The figure also suggests that the nuclear separation between two galaxies in a pair  should be used with caution as it does not   vary linearly along the merger sequence.
The statistical results based on   $r_{p}$ may depend on the number of galaxies in each stages.

\subsection{Spatially resolved \SDSFR{}}
\label{sec_resolved}
\begin{figure*}
	\begin{center}
		\subfigure[]{\label{fig_Resolved_dSFR_hist_full}\includegraphics[scale=0.55]{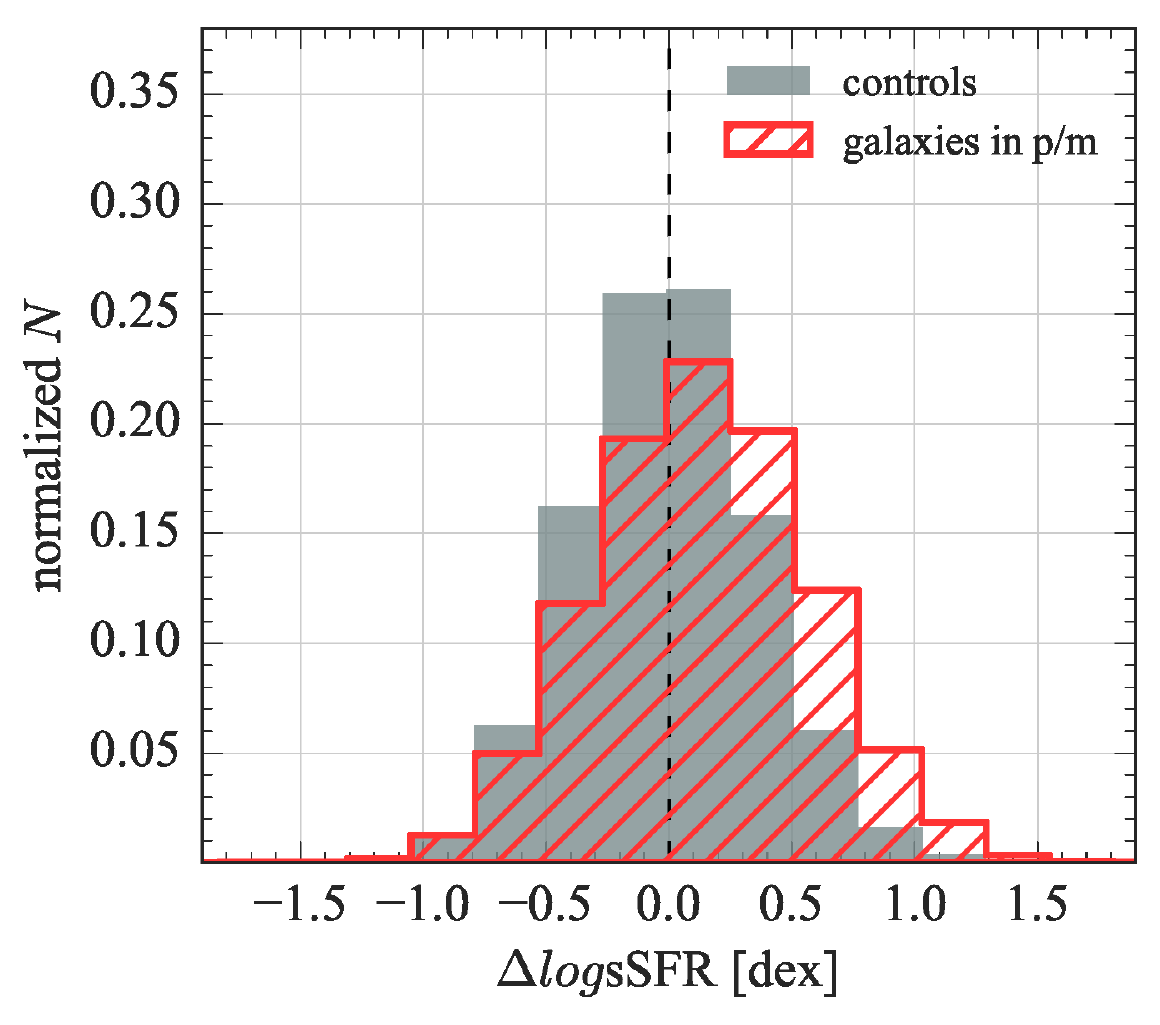}}
		\subfigure[]{\label{fig_Resolved_All_full}\includegraphics[scale=0.55]{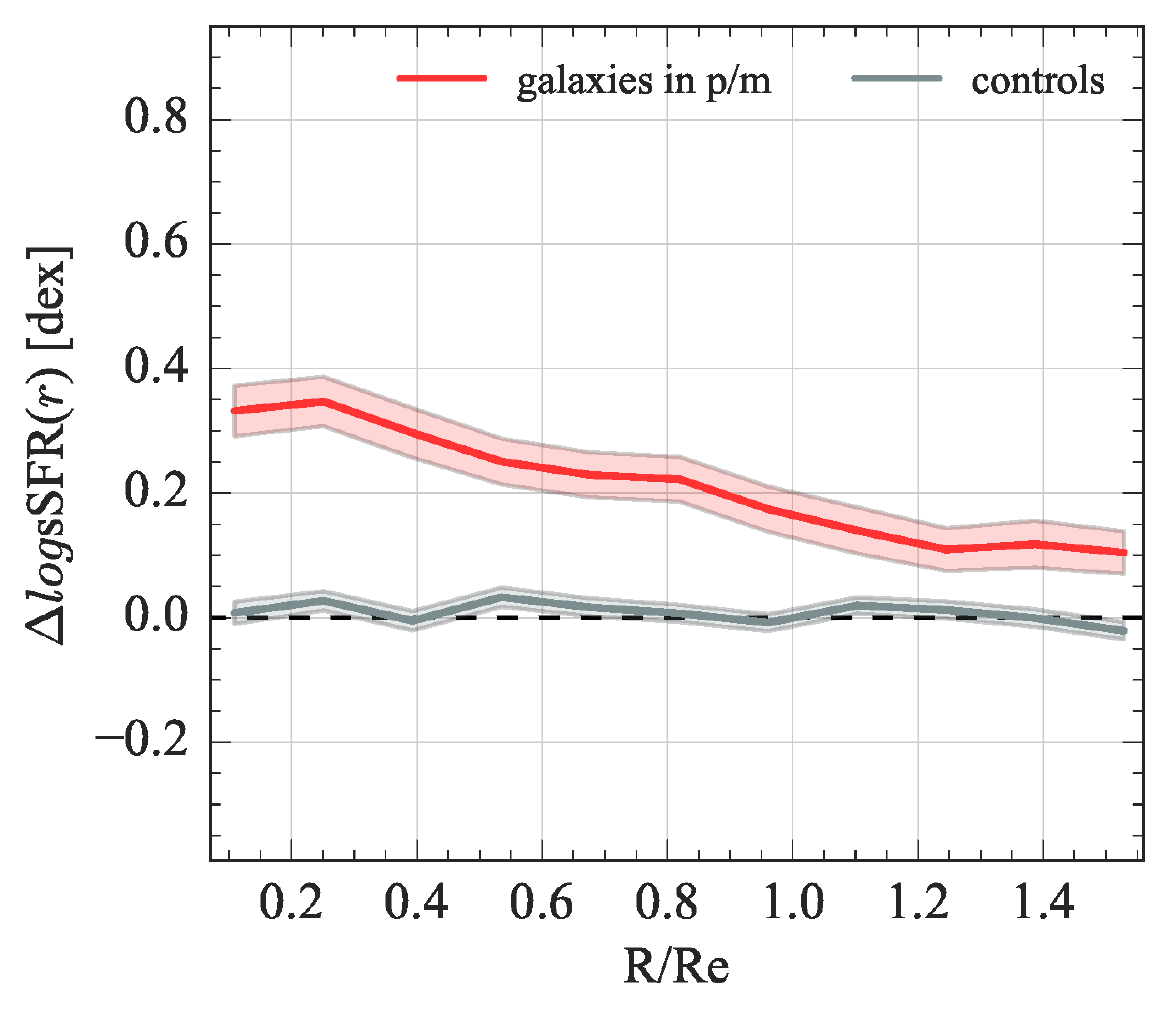}}
		\subfigure[]{\label{fig_Resolved_check_control}\includegraphics[scale=0.55]{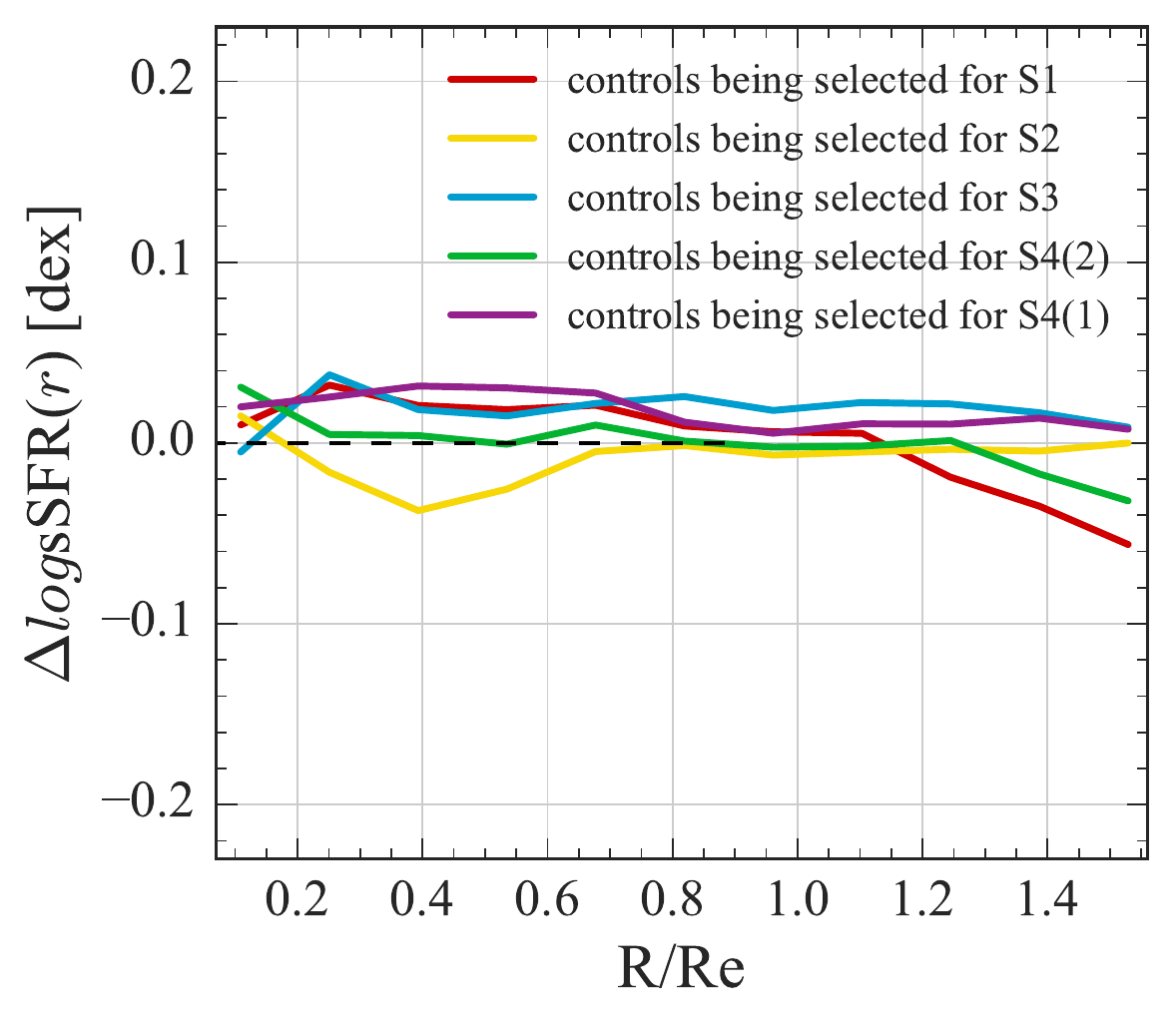}}
		\subfigure[]{\label{fig_all_s}\includegraphics[scale=0.55]{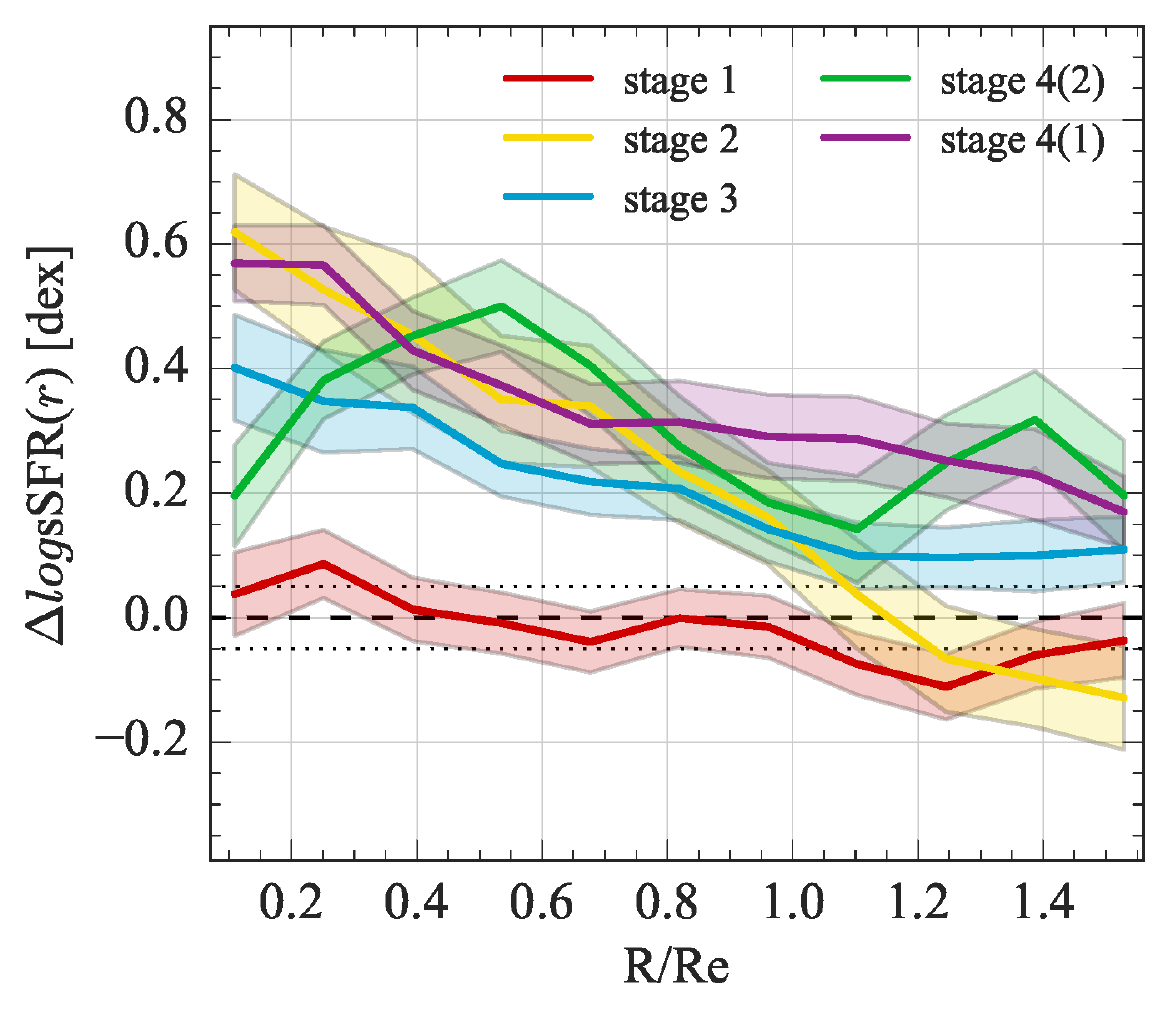}} 
	\end{center}
	 %\vspace{-10pt}
	\caption{(a) Distribution of spaxel-wise $\Delta$$log$sSFR  of the galaxies in \emph{p/m} (hatched histogram) and the pool of control galaxies (solid histogram). The vertical line indicates the zero line. (b) Radial \SDSFR{} of the galaxies in \emph{p/m} (red) and the pool of control galaxies (gray). The horizontal line indicates the zero line. The shaded areas correspond to the standard error of the mean.  (c) Median \SDSFR{} profiles of the controls being selected for Stage 1 (red), Stage 2 (yellow), Stage 3 (blue), Stage 4(2) (green), and Stage 1(1)  (purple). The plot indicates the systematic variations of   \SDSFR{} profiles of galaxies in \emph{p/m} (note that the range of $y$-axis of this figure is narrower than other panels).  (d) Radial \SDSFR{} of galaxies in \emph{p/m} in different merger stage. Red, yellow, blue, green, and purple lines represent Stages 1, 2, 3, 4(2), and 4(1) respectively. The shaded areas represent  the standard error of the mean.  Two dotted lines at $\pm$ 0.05 dex are plotted to indicate the impact of intrinsic signatures in the control sample. The systematic variations (panel c) have been subtracted from the radial profiles.
	}
	\label{fig_resolved_all}
\end{figure*}

%\subsubsection{\SDSFR{} Distribution and Radial Profiles}
\label{sec_spatial}
We first compare the local sSFR  of  galaxies in \emph{p/m} and the controls by means of their spaxel-by-spaxel distributions.
This is shown in Figure \ref{fig_Resolved_dSFR_hist_full}, where  the distribution of  the local $\Delta$$log$sSFR of  the galaxies in \emph{p/m} is shown by hatched histogram and that of  the galaxies in the control pool by solid histogram.
Again, the distribution of controls indicates the intrinsic scatter of $\Delta$$log$sSFR.
As is clearly seen in the plot, galaxies in \emph{p/m} have higher $\Delta$$log$sSFR than the controls.
The  standard deviation of the $\Delta$$log$sSFR values is 0.37 dex for the galaxies in the control pool and 0.44 dex for the galaxies in \emph{p/m}.
The median $\Delta$$log$sSFR of galaxies in \emph{p/m} and controls are 0.14 dex and 0.01 dex, respectively.

Turning now to the  radial  distribution of \SDSFR{}, Figure \ref{fig_Resolved_All_full} shows the median radial  distribution of the \SDSFR{} for the galaxies in \emph{p/m} (red) and galaxies in the control pool (gray).
The shaded regions represent the error on the mean.
%For each $R_\mathrm{e}$ bin, the shaded regions represent the error on the median gradients (solid color lines), which take into account the number of spaxels contributing at each radius. 
The horizontal dashed line indicates zero enhancement.
As might reasonably be expected, galaxies from the control pool exhibit a  flat \SDSFR{} profile around zero,  confirming  once more that the method we have taken to compute the \SDSFR{}  is valid.
In contrast, the  radial distribution for  galaxies in \emph{p/m} reaches \SDSFR{} of $\sim$ 0.3 dex in the centers, then decreases radially to a value of $\sim$ 0.1 dex at the edge of the radius we explore.
The  profile suggests that, when considering all the merger populations as a whole, \emph{local sSFR is enhanced   not only in the central region but also in the disk during galaxy interactions}, in agreement with  recent simulations \citep[e.g.,][]{Tey10,Per11,Hop13,Pow13,Ren15,Ren16,Sil17}  and observations \citep[e.g.,][]{Bar15a,Cor17c}.

To gain more insight on when the enhanced local sSFR  is actually taking place during galaxy interactions, we construct the radial  \SDSFR{} distributions  for different merger stages.
We first  estimate the systematic variation of \SDSFR{} for different merger stages in Figure \ref{fig_Resolved_check_control} (note that the range of $y$-axis of this figure is narrower than other panels) by computing the median radial  \SDSFR{}  distributions  of the  \emph{control galaxies that have been used}  for individual stages.
The systematic variations are in the range of $\pm$ 0.05 dex.
Figure \ref{fig_all_s} presents the distributions of radial \SDSFR{}  for different merger stages.
Although the systematic variations in Figure \ref{fig_Resolved_check_control} have been subtracted from this plot,  two dotted lines at $\pm$ 0.05 dex are plotted to indicate the impact of  intrinsic signatures in the control sample.
At \textit{S1}, the radial \SDSFR{} distribution has a relatively flat profile within $\pm$ 0.1 dex, suggesting that galaxy interactions have almost no impact on star formation during the incoming phase.
At the first pericenter passage (\textit{S2}), the \SDSFR{} distribution  presents a relatively steep profile that is not seen in other stages.
The \SDSFR{} decreases from $\sim$ 0.6 dex at the innermost region to $\sim$ -0.15 dex at 1.5 $R_{e}$.
%The most important point to take from these plots is that the
Later on, extended star formation enhancements are observed at all stages after the first pericenter passage (\textit{S3}, \textit{S4(2)}, and \textit{S4(1)}), but the magnitudes and the profiles are somewhat different among these stages.
At \textit{S3} and \textit{S4(1)}, overall, \SDSFR{} values increase radially inward.
In spite of the similar trend of variation, the median \SDSFR{} profile of galaxies at \textit{S4(1)} lies above that of galaxies at \textit{S3} at all radii by $\sim$ 0.1 -- 0.2 dex. 
Moving to \textit{S4(2)} galaxies, we see that  
the  profile  shows no obvious radial dependence, fluctuating between $\sim$ 0.2  and 0.6 dex.
This fluctuating behavior is presumably related to their chaotic morphologies (Figure \ref{fig_stages}(d)(e)).

We have checked  whether the radial   \SDSFR{} distributions  depend on redshift; for instance, at higher $z$, because of the resolution,    profiles may tend to appear flatter than at lower $z$.
This is done by reproducing Figure \ref{fig_all_s} but now using galaxies with $z$ $<$ 0.03 (mean $z$ of our sample) and $z$ $>$ 0.03, separately.
All of the observed features in  Figure \ref{fig_all_s}  are clearly present in the sub-sample plots.
Therefore, we stress that redshift (resolution) has negligible effect on our conclusions.

In summary, we find that (1) the highest,  interaction-triggered star formation, in both the global and the local sense, occurs in the  \textit{S4(1)}, where the nuclei of two galaxies have merged,  (2) while the values of global \GDSFR{} for   each of the different stages  are not dramatically different (Figure \ref{fig_all_g}), diverse  variations  in  radial \SDSFR{} profile are observed along the merger sequence, and (3) interaction-triggered star formation is not restricted to the central region of a galaxy.

\

\section{Discussion} % and comparison with other results from the literature
\label{sec_discussion}

\subsection{Evolutionary Scheme}
\label{sec_discuss_distribution}

Our results  demonstrate that galaxy interactions trigger  centrally peaked star formation since the first pericenter passage (\textit{S2}) and extended star formation throughout the interaction phase until the final post-coelescence (\textit{S3}--\textit{S4}).
Our  results are in global agreement with recent numerical simulations \citep{Per11,Sil17} and  results based on  single-fiber SDSS measurements \citep{Ell13}.
Here we discuss the possible origin of the varying radial \SDSFR{} profiles among the merger stages.

Galaxy interactions can cause gas that has previously settled in the outskirts to be funneled  towards the central region, resulting in an enhancement of star formation in this region \citep{Bar91}.
Our results imply that such gas inflows become efficient at the first pericenter passage (\textit{S2}), which is consistent with many simulations. 
The boost of gas inflow rate  can trigger the steep  radial distribution of \SDSFR{} and significant enhancement of star formation  in the central regions \citep[e.g.,][]{Mor15}.
For equal-mass gas-rich disk-disk interactions, the average gas inflow rate is $\sim$ 2 M$_{\sun}$ yr$^{-1}$ over $\sim$ 200 Myr \citep{Tor12}, and the mass that is expected to be deposited into the inner region between the first passage to the coalescence phase (i.e., \textit{S2} and \textit{S3}) is therefore $\sim$ 4 $\times$ 10$^{8}$ M$_{\sun}$, accounting for $\sim$ 20\% of the gas for a Milky Way-like galaxy (note that the adopted time duration of gas inflow is shorter than the time between the first pericenter passage and the coalescence phase because the inflow rate decreases rapidly after the first apocenter passage).
Therefore, although galaxy interaction drives gas inwards, the galactic disk still contains a significant amount of gas.

After the first passage, star formation is  enhanced within both the inner regions and at larger galactocentric radii.
Several possible mechanisms may contribute to the extended star formation.
First of all, part of the extended star formation activity could be  fed by gas accretion from the companion  after the pericenter, but the
 level of enhancement depends on several factors such as geometry and gas fraction  \citep{Per11,Sil17}.
Secondly, recent simulations show that  the outer regions of interacting systems receives a significant fraction of gas  coming from the inner regions of the disks as the arms get opened and distorted by tides  \citep{Per11}. This gas subsequently feeds the star formation in the outer regions.

Furthermore, the  extended star formation activity could be attributed to the change of  the properties of the existing gas.
Galaxy interactions  produce convergent flows and shocks throughout the galaxy  \citep{Tey10,{Pow13}}. 
These flows and shocks can alter  gas in two manners. 
One is that the flows and shocks induce phase transitions (e.g., HI $\rightarrow$ H2; \citealt{Mor19})  throughout the galaxy.
Indeed, a number of observations have found such an increase in the molecular gas fraction in galaxies in \emph{p/m} compared to the isolated galaxies (\citealt{Vio18}; \citealt{Pan18}; Sargent et al. in prep.). 
The other possibility is that the gas turbulence increases through the flows and shocks, so that gas clouds become more massive and denser than in isolated galaxies, and the free-fall time becomes shorter. 
In this case, mergers can convert gas into stars faster than isolated galaxies with similar gas surface densities, i.e. higher star formation efficiency.
This scenario is consistent with the  observational facts that interacting galaxies show higher molecular cloud mass, higher velocity dispersion of gas, and higher star formation efficiency compared to galaxies in isolation \citep[e.g.,][]{Elm95a,Elm95b,Wil03,Elm00,Str05,Her11,Hug13,Mic16}.
Finally,  the first pericenter passage can trigger the formation of spiral arms and large-scale filament structures, where gas can be efficiently compressed to form stars in the disk regions \citep[e.g.,][]{Elm06,Pet17,Esp18}.

In any case, it is worth pointing out that while most  existing   simulations and observations  of  galaxy  interactions,  in  particular those studying the spatial extent of interaction-triggered star formation activity, focus on major mergers, our analysis, which consists of a wide range of mass ratios, also suggests the existence of  extended interaction-triggered star formation.
Therefore,  extended star formation is probably not restricted to equal-mass systems like the Antennae.

While a proposed star formation quenching mechanism  is galaxy mergers, the permanent star formation quenching that might be expected to  result from mergers has not occurred throughout the four merger phases that we explore \citep[see also][]{Tho18}.
\citet{Spa17} investigated how  morphological  transformations and quenching occur in galaxy mergers using cosmological simulations.
They found that star formation of post-mergers (merger remnants) is not necessarily quenched unless the AGN feedback is sufficiently strong  (see also \citealt{San18}).
Alternatively, both  major and minor post-mergers can potentially have star-forming disks if there is enough gas available at the coalescence phase \citep{Kau93,Rob06}. 
This  is supported by observations of the cold gas (disks) in  post-mergers \citep{Ued14,Ell15}.

\subsection{Comparison with Previous Studies}
\citet{Bar15a}  carried out the first statistical study of the impact of the merger event on the spatial star formation distribution using 103 galaxies in \emph{p/m} and 80 controls from the  CALIFA  IFS survey.
They found a moderate enhancement in the sSFR ($\times$ 2 -- 3) in the central region of galaxies in \emph{p/m} (see also \citealt{Scu12} and  \citealt{Ell13} for  moderate central SFR triggering). 
We find a similar value for our sample (Figure \ref{fig_Resolved_All_full}). 
In \citet{Bar15a}, the sSFR is similar  to  or  moderately  suppressed  in  comparison  to  the  control sample in the outer regions,  while we obtain a higher level of 0.1 -- 0.2 dex  (Figure \ref{fig_Resolved_All_full}).
This discrepancy could   originate from a different sample selection, e.g., \citet{Bar15a} adopted a much wider separation criterion than this work ($\sim$160 kpc versus $\sim$71 kpc), and therefore their sample potentially contains more galaxies with lower level of triggered star formation.
In spite of the different sample selection criteria,  both studies have led to a fairly similar understanding that  galaxy interactions have larger impact on central star formation than on disk  star formation.
This concept was indirectly probed through  single-fiber observations  by \citet{Ell13}, and has now been directly established  by the IFS surveys.

Our results show that interaction-triggered star formation is more concentrated in the inner part of the systems at the first pericenter passage (\textit{S2}); but in later stages, the interaction-triggered star formation is more extended (\textit{S3} -- \textit{S4}).
The scenario is  in broad agreement with the results obtained previously by  \citet{Ell13}. 
However, by analyzing IFS data of  three luminous infrared galaxy (LIRG; $L_\mathrm{IR}$ $>$ 10$^{11}$ $L_{\sun}$) mergers  from the CALIFA survey,  an opposite trend was reported by \citet{Cor17c}. 
Nonetheless, it is interesting to note that the non-LIRG system (the Mice), which corresponds to our \textit{S2}, in  \citet{Cor17c} shows suppressed star formation in the disk regions,  just like  our galaxies at  \textit{S2}.
The apparent discrepancy between normal galaxies in \emph{p/m} (sample in \citealt{Ell13}, the Mice system in \citealt{Cor17c}, and galaxies in this work) and high-luminosity mergers  (the three LIRG mergers in \citealp{Cor17c})  suggests that in addition to merger stage, other physical mechanisms, such as the  gas fraction (gas mass with respect to stellar mass), gas properties (e.g., dense gas fraction; dense gas mass with respect to total gas mass),  orbital characteristics, and properties of progenitors (e.g., gas rich or not),  also play a  role in triggering star formation \citep[e.g.,][]{Dim07,Dim08,Cox08,Bus18}.

\subsection{Effect of Fiber Coverage on Interaction-triggered Global (s)SFR}
\label{sec_aperture}
MaNGA data allow us to evaluate the potential effect of aperture size  on \emph{global} interaction-triggered star formation. 
What is clear from our analysis is  that the magnitude of interaction-triggered star formation  varies with radius, with the largest enhancement in the central regions.
As such,  the  physical  coverage of a fixed angular fiber, such as  used in  the traditional SDSS observations, and how  the aperture correction of SFR is being made may affect the derived interaction-triggered star formation rate \citep[see also][]{Ell13,Patt13,Bar15a}.

\begin{figure}
	\centering
	\includegraphics[scale=0.45]{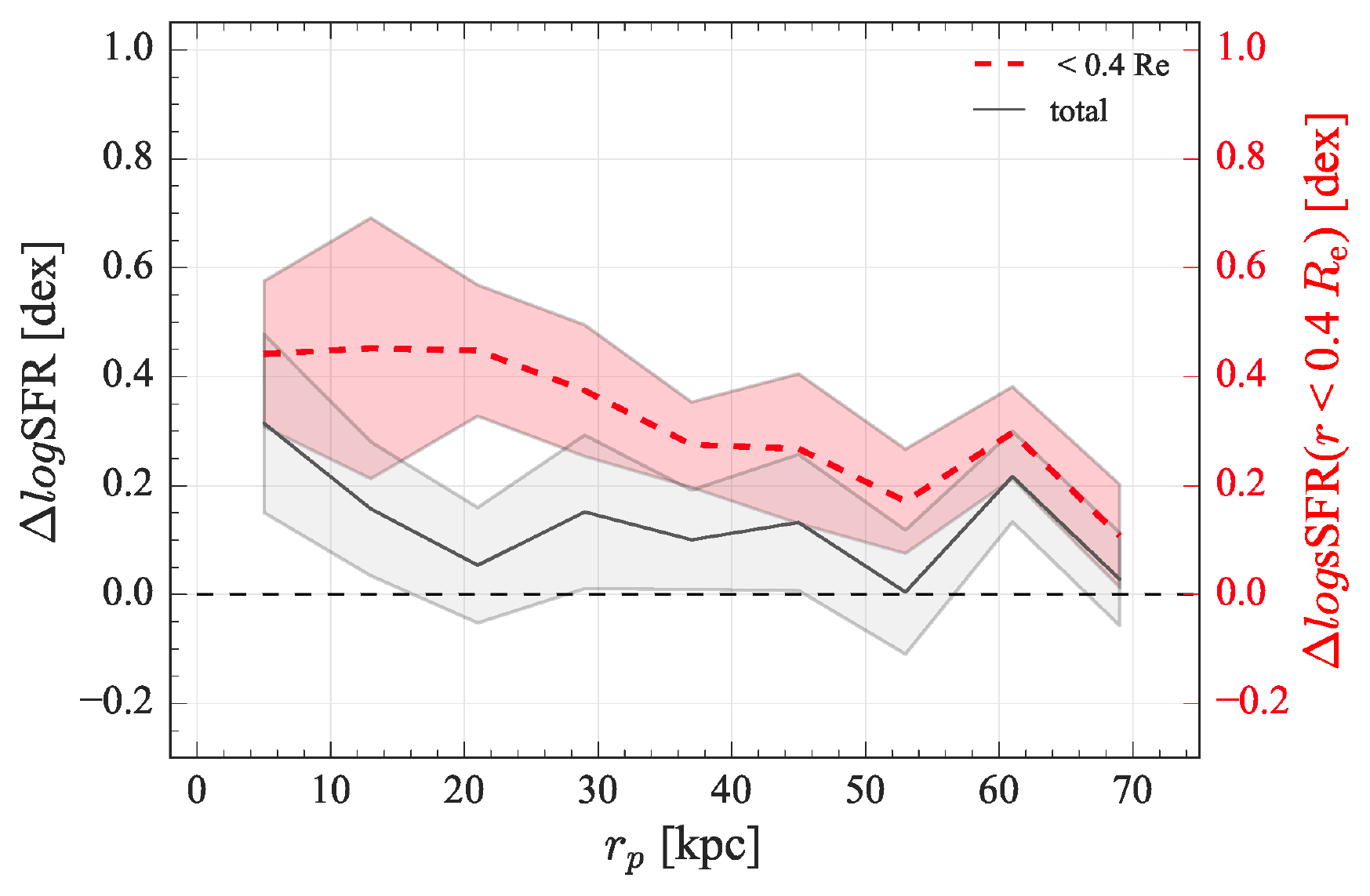}
	\caption
	{Star formation enhancement versus  projected separation. The solid and dashed lines represent median enhancement of global SFR and sSFR within the central 0.4$R_{e}$, respectively. The shaded area indicates the standard error of the mean. The horizontal dashed line indicates the zero line. Note that the $y$-axis of the  dashed and  solid lines are different due to the different  methods of analysis (see text for details). The  $y$-axis of the black solid line is \GDSFR{} (black axis label at the left; same as in Figure \ref{fig_Projsep_large}), while the  $y$-axis of the red dashed line is $\Delta$$log$sSFR($r$ $<$ 0.4$R_\mathrm{e}$) (red axis label at the right).
	%The general trend of total \GDSFR{} versus $r_{p}$ is preserved for the central $\Delta$sSFR($<$ 0.4$R_{e}$), but the level of enhancement is larger for the $\Delta$sSFR($<$ 0.4$R_{e}$) due to the centrally-peaked \SDSFR{} profiles in Figure \ref{fig_all_s}.
	}
	\label{fig_SFR_scale}
\end{figure}

For each galaxy, we generate mock single fiber measurements by summing up spaxel-wise SFR and $M_{\ast}$ within 0.4$R_{e}$, and then calculate the $\Delta$$log$sSFR($r$$<$ 0.4$R_{e}$) in comparison to its control galaxies selected in Section \ref{sec_global_offset}.
The median $\Delta$$log$sSFR($r$$<$ 0.4$R_{e}$) as a function of $r_{p}$ is shown in Figure \ref{fig_SFR_scale} with a  dashed line, while the result based on total SFR (Figure \ref{fig_Projsep_large}) is  overlaid in the figure with a black solid line.
 It should be  emphasized that the $y$-axis  represents the excess in SFR for the global measurements (solid line) and the excess in sSFR for  the central measurements (dashed line)  due to the different  methods of analysis. 
Despite the difference, the values of the global \GDSFR{} in Figure \ref{fig_Projsep_large} is approximately equivalent to their excess in the global sSFR as the stellar mass has been matched between galaxies in \emph{p/m} and controls (\S\ref{sec_global_offset}). 

As  expected, Figure \ref{fig_SFR_scale} suggests that the central star formation is more elevated than the global star formation because centrally-concentrated star formation enhancement is
common (Figure \ref{fig_resolved_all}).
Moreover,  $\Delta$$log$sSFR($r$$<$ 0.4$R_{e}$) shows a significant excess  in  the  galaxies in \emph{p/m}    relative  to  their controls out to $\sim$ 70 kpc, while the global \GDSFR{}  is significantly enhanced  only when the pairs are at $r_{p}$ $<$ 20 kpc.
Thus, caution is needed when quoting the degree of interaction-triggered star formation  obtained from single-fiber observations as they  tend to probe the regions with the strongest enhancements \citep[see also][]{Lam03,Patt13,Ell13,Bar15a}.

\subsection{Caveats}
\label{sec_caveats}

\subsubsection{Stage Classification}
Here we discuss  any possible limitations and caveats to our classification scheme.
First of all, one could imagine a possible bias against higher $z$ for galaxies without morphology  distortion/asymmetries due to a sensitivity limit.
While the galaxies are far from their companions, the presence of  morphology distortion/asymmetries is used  to distinguish \textit{S1} and \textit{S3} (all of them are identified as galaxies in \emph{p/m} via spectroscopic data),  however, the  morphology distortions/asymmetries in the outskirts  could be faint  compared to the detection limits.
In other words, the visibility of these features  depends on the sensitivity.
In order to evaluate the potential bias of the visual inspection, we check the $z$ distributions of galaxies in \textit{S1} and \textit{S3}.
The median $z$ of galaxies in   \textit{S1} and \textit{S3} are almost identical, 0.0281 and 0.0286 respectively. 
A Kolmogorov-Smirnov test confirms that with a $p$-value of 0.24, the $z$ distributions for \textit{S1} and \textit{S3} are statistically the same. 
Accordingly, we find no evidence for a strong bias due to sensitivity for \textit{S1} and \textit{S3}.
Moreover, the evidences of interaction in the \textit{S4} (e.g., shell structures at the outskirts) would be difficult to detect as the redshift increases. 
Although we find no difference in the median $z$ of  \textit{S4} (0.0296) and that of other stages,  we should keep in mind that the depth of images play a critical role in identifying galaxies in \emph{p/m} \citep[e.g.,][]{Ell19}.

Secondly, it is possible that some systems would not proceed through the assumed merger scenario, e.g.,  two  galaxies encounter  but  separate  at  a  later  time and  do not ever return (i.e., flybys)  or two galaxies directly merge during the first encounter (as seen in some simulations).
Observationally, it is challenging to  account for the  effects  of  flybys and  direct mergers  directly due to the unknown orbits  of two approaching galaxies.

It is interesting to note that galaxy  flybys are not rare, the  galaxy-pair  counts in a large-scale survey (e.g., SDSS) at $z$ $<$ 3  are   contaminated by flybys at least at a 20 -- 30\% level, as estimated by the simulations of \citet{Sin12}.
Galaxy flybys are  capable of causing  perturbations  \citep{Wei09,Paw11,Kim14,Lan14,Zan18}; therefore,  effects on star formation are expected. 
Flyby interactions can be present in \textit{S1} -- \textit{S3} (pre-flybys, ongoing flybys, and post-flybys, respectively).
However, even though the flybys can exert effects on galaxy evolution, perhaps  not all of them are able to  generate the necessary (observable) features to be identified as in \textit{S2} and \textit{S3} according to our classification  scheme.

Direct mergers are observed in cosmological simulations, and even occur in major mergers \citep{Bus18}.
Therefore, some \textit{S2} systems, but perhaps not many, may in fact be in their final coalescence phase and would not experience extra apocentric passage.
On the other hand, some systems at  \textit{S4} may have experienced a direct merger without extra pericenter passages.

Finally, even though the galactic disks of our controls are rather symmetric,  we cannot fully rule out the possibility that they experienced past external perturbations  such as flybys or minor mergers,    which may or may not have altered their star formation activity.

\subsubsection{Merger Configurations}
\label{sec_caveat_stage}
Our observational data include galaxies that could have any merger configuration.
There are many  factors regarding merger configurations that can affect the magnitude of interaction-triggered star formation, such as the  properties of the companion \citep{Hwa10,Xu12,Cao16,Sil18}, mass ratio  \citep{Cox08,Bus18}, cold gas reservoir \citep{Scu15,Vio18,Pan18}, and the encounter geometry of the two interacting galaxies \citep{Dim07,Dim08}.
Moreover, it has been shown in many observations that while the primary and secondary galaxies in a major merger exhibit symmetry in their response to the tidal interaction, the star formation activity in the less massive member in the  minor merger suffers a more dramatic impact than that of its companion during the interactions \citep[e.g.,][]{Woo07,Sil18}.
The current sample size of MaNGA is not sufficiently large to  enable   further statistical analysis of the dependence of triggering star formation on merger configurations.
The full MaNGA sample -- in which the number of galaxies in \emph{p/m} is expected to be  at least doubled (similar for the control sample) -- will provide further constraints on the evolutionary scheme of interaction-triggered star formation and their effects on galaxy evolution.

\subsubsection{Internal Structures of Disk Galaxies}

Our star-forming sample shows a significant diversity in their internal structures, such as the existence of a bar and a varing number of spiral arms and pitch angles.
This may introduce systematic uncertainties in the radial \SDSFR{} distribution  if the local star formation  of galaxies is intrinsically correlated with the internal structures \citep[e.g.,][]{Gon16,Har17a,Har17b}.
For instance, we may have compared a galaxy with actively star-forming spiral arms with a galaxy with a relatively smooth,  moderately star-forming disk. 
The comparison might be reinforced  by  matching the control sample in  detailed  morphology \citep[e.g., T-type morphology;][]{Fis19}, but this would require a significant increase in the pool of controls.
Besides, the formation of galactic internal structures  may be closely tied with galaxy interactions \citep{Elm06,Lan14,Lok16,Pet17,Esp18}.

\section{Summary}
\label{sec_summary}
In this paper,  we present  an empirical picture of the evolution of the star formation distribution in interacting galaxies and mergers. 
We analyze the global and local star formation of 205 star-forming (log(sSFR/yr$^{-1}$) $>$ -11) galaxies    in pairs/mergers observed by SDSS-IV MaNGA.
We consider an interacting galaxy to be one which has a spectroscopic companion or has experienced a significant tidal force, showing morphology distortion or ongoing interaction with a companion galaxy (\S\ref{sec_pair_select_all}). 

Merger stage is identified through visual examination of each interacting galaxy selected (\S\ref{sec_merger_stage} and Figure \ref{fig_stages}),  \textit{Stage 1 (S1)}: before the first pericenter passage, without morphology distortion, \textit{Stage 2 (S2)}: at   the first pericenter passage, with a close companion and tidal bridges,  \textit{Stage 3 (S3)}: after the  the first pericenter passage, around the apocenter, and with morphological  distortion, and  \textit{Stage 4 (S4)}: in the coalescence phase. For \textit{S4}. 
We further classify \textit{S4} into \textit{S4(2)} and \textit{S4(1)} according to the number of visible cores.

To quantify the interaction-triggered star formation activity,  we identify a pool of star-forming control galaxies ($\sim$ 1350) from MaNGA (\S\ref{sec_control}).
For each galaxy, we  select control galaxies from the pool satisfying the matching conditions in global galaxy properties.  
Then the global SFR and local sSFR  with respect to the controls: \GDSFR{} and radial \SDSFR{} distribution, are computed (\S\ref{sec_offset}).
The main results are summarized as follows.

\begin{itemize}
	\item Using the computed values of \GDSFR{}, we confirm  previous results that, statistically,  global SFR is enhanced by a limited level  during galaxy interactions. Moreover, \GDSFR{} increases with decreasing nuclear separation ($r_{p}$) between two galaxies in a pair. However, we find that the trend could be sensitive to the fraction of galaxies in different merger stages because   $r_{p}$ does not vary linearly along the merging sequence   (\S\ref{sec_global} and Figures \ref{fig_global_all}, \ref{fig_Projsep_large}). 
	
	\item While the values of global \GDSFR{} for   each of the different stages  are not significantly different, there is considerable variation in the radial \SDSFR{} profiles across the merger stages.
	 Statistically, galaxy interactions have no impact on local star formation during the incoming phase  (\textit{S1}). Right after the first pericenter passage (\textit{S2}), galaxy interactions  produce enhanced star formation in the center, and suppressed star formation in the outskirts, resulting in a steep profile in the radial \SDSFR{} distribution.
	 After the first pericenter passage (\textit{S3}), star formation  is  enhanced all the way out to the galactocentric radius we explore (1.5 $R_{e}$ $\approx$ 6.7 kpc). The radial  \SDSFR{} profile at \textit{S4(2)} shows no radial dependence due to their chaotic morphologies.
	 Finally, the highest,  interaction-triggered star formation, in both the global and the local sense, occurs in the  \textit{S4(1)}, where the nuclei of two galaxies have merged (\S\ref{sec_resolved} and Figure \ref{fig_resolved_all}).
	
	\item Our results suggest that interaction-triggered star formation is not restricted to the central region of a galaxy (although the enhancement is indeed centrally peaked).
	 In addition to the well-known gas inflow, other mechanisms, such as  gas accretion from the companions,  interaction-triggered gas migration, or interaction-triggered internal structure formation, and an increase in gas turbulence can control the  level and distribution of the  star formation in  galaxies in pairs/mergers (\S\ref{sec_discuss_distribution}). 
	 
	 \item Our analysis clearly shows that the magnitude of interaction-triggered star formation varies with galactocentric radius. As such, the physical coverage of a fixed angular fiber, such as the traditional SDSS observations, and how  the aperture correction of SFR is  made may affect the derived interaction-triggered star formation rate. Specifically, the degree of enhancement decreases as physical fiber coverage increases due to dilution of triggered star formation by disk regions (\S\ref{sec_aperture} and Figure  \ref{fig_SFR_scale}).
	 
\end{itemize}

Our understanding of the mechanisms that trigger star formation during galaxy interactions is far from  complete. 
In future work, we intend to investigate (1) the dependence of the interaction-triggered star formation on merger configurations (mass ratio, wet mergers and dry mergers), (2) the interaction-triggered changes of  metallicity of galaxies  (Barrera-Ballesteros et al. in prep.), using our MaNGA data, and (3) the  evolution of molecular gas (i.e., stellar nurseries) properties along the merger sequence, using   ALMA data.
The explorations of a wide range in parameter space and interstellar medium (ISM)  properties  is  required to constrain the whole picture of interaction-triggered star formation.

\section*{Acknowledgements}

We thank the anonymous referee for useful comments which
	improved this paper.  The authors also thank Sara Ellison and Mallory Thorp for many useful and enjoyable discussions regarding galaxy mergers and Gaoxiang Jin for discussions on merger stage classification.
The work is supported by the Academia Sinica under the Career Development Award CDA-107-M03 and the Ministry of Science \& Technology of Taiwan under the grant MOST 107-2119-M-001-024-.
PBT acknowledges partial support from UNAB research grant 2019.
MB acknowledges the FONDECYT regular grant 1170618.
RR thanks FAPERGS, CNPq and CAPES.
J.H.K. acknowledges financial support from the European Union’s Horizon 2020 research and innovation programme under Marie Sk\l odowska-Curie grant agreement No 721463 to the SUNDIAL ITN network, from the Spanish Ministry of Economy and Competitiveness (MINECO) under grant number AYA2016-76219-P, from the Fundaci\'on BBVA under its 2017 programme of assistance to scientific research groups, for the project ``Using machine-learning techniques to drag galaxies from the noise in deep imaging'', and from the Leverhulme Trust through the award of a Visiting Professorship at LJMU.

This project makes use of the MaNGA-Pipe3D dataproducts. We thank the IA-UNAM MaNGA team for creating this catalogue, and the ConaCyt-180125 project for supporting them.
We also thank  the  MPA/JHU  teams  for  making their catalogs publicly available. 

Funding for the Sloan Digital Sky Survey IV has been provided by the Alfred P. Sloan Foundation, the U.S. Department of Energy Office of Science, and the Participating Institutions. SDSS acknowledges support and resources from the Center for High-Performance Computing at the University of Utah. The SDSS web site is www.sdss.org.

SDSS is managed by the Astrophysical Research Consortium for the Participating Institutions of the SDSS Collaboration including the Brazilian Participation Group, the Carnegie Institution for Science, Carnegie Mellon University, the Chilean Participation Group, the French Participation Group, Harvard-Smithsonian Center for Astrophysics, Instituto de Astrofísica de Canarias, The Johns Hopkins University, Kavli Institute for the Physics and Mathematics of the Universe (IPMU) / University of Tokyo, the Korean Participation Group, Lawrence Berkeley National Laboratory, Leibniz Institut f\"ur Astrophysik Potsdam (AIP), Max-Planck-Institut f\"ur Astronomie (MPIA Heidelberg), Max-Planck-Institut f\"ur Astrophysik (MPA Garching), Max-Planck-Institut f\"ur Extraterrestrische Physik (MPE), National Astronomical Observatories of China, New Mexico State University, New York University, University of Notre Dame, Observatório Nacional / MCTI, The Ohio State University, Pennsylvania State University, Shanghai Astronomical Observatory, United Kingdom Participation Group, Universidad Nacional Autónoma de México, University of Arizona, University of Colorado Boulder, University of Oxford, University of Portsmouth, University of Utah, University of Virginia, University of Washington, University of Wisconsin, Vanderbilt University, and Yale University.

\end{document}